# UNIVERSAL MACHINE LEARNING INTERATOMIC POTENTIALS ARE READY FOR SOLID ION CONDUCTORS

**Hongwei Du[1,2,3], Jian Hui[1,2,3]*, Lanting Zhang[1,2,3]*, Hong Wang[1,2,3]***

[1] School of Materials Science and Engineering, Shanghai Jiao Tong University, Shanghai 200240, China.

[2] Zhangjiang Institute for Advanced Study, Shanghai Jiao Tong University, Shanghai 201203, China.

[3] Materials Genome Initiative Center, Shanghai Jiao Tong University, Shanghai 200240, China.

**ABSTRACT**

With the rapid development of energy storage technology, high-performance solid-state electrolytes (SSEs) have become key materials for next-generation lithium-ion batteries. These materials need to possess high ionic conductivity, excellent electrochemical stability, and good mechanical properties to meet the demands of electric vehicles and portable electronic devices. However, traditional design methods, such as density functional theory (DFT) calculations and empirical force field simulations, face challenges such as high computational cost, poor scalability, and insufficient accuracy across different material systems. In this context, universal machine learning interatomic potentials (uMLIPs) have attracted significant attention due to their high efficiency and near-DFT-level prediction accuracy. This study develops a systematic benchmark framework for SSEs to evaluate the performance of state-of-the-art uMLIP models in terms of energy, forces, thermodynamic properties, elastic moduli, and lithium-ion diffusion behavior. Through this framework, we comprehensively assessed the applicability of advanced models such as MatterSim, MACE, SevenNet, CHGNet, M3GNet, and ORBFF in complex material systems. The results show that MatterSim performs best in almost all metrics, particularly excelling in computational accuracy and physical consistency in complex material systems compared to other models. In contrast, other models exhibit significant deviations in force calculations and thermodynamic property calculation due to a lack of energy consistency or insufficient training data coverage. Further analysis reveals that MatterSim achieves excellent agreement with reference values in lithium-ion diffusivity calculations, especially at room temperature. Additionally, based on MatterSim, studies on two typical SSEs, $Li_3YCl_6$ and $Li_6PS_5Cl$, uncover the mechanisms by which crystal structure, anion disorder levels, and Na/Li arrangements influence ionic conductivity. The study demonstrates that appropriate S/Cl disorder levels and optimized Na/Li arrangements can significantly enhance the connectivity of lithium-ion diffusion pathways, thereby improving the overall ionic transport performance of the materials. In summary, this study not only validates the superiority of MatterSim in complex material systems but also provides crucial theoretical support for the design and optimization of SSEs, while significantly reducing the cost of discovering and evaluating new materials. Furthermore, the benchmark framework developed in this study offers a standardized tool for assessing the performance of novel uMLIP models in the future, holding significant academic and practical value.

## 1. Introduction

With the rapid advancement of energy storage technologies, high-performance solid-state electrolytes (SSEs) have emerged as critical materials for next-generation lithium-ion batteries[1-4]. These materials are required to exhibit high ionic conductivity, excellent electrochemical stability, and robust mechanical strength to meet the demands of modern applications such as electric vehicles and portable electronic devices[5-10]. However, traditional design and screening methods, such as density functional theory (DFT) calculations and empirical force field simulations[11-15], often face challenges such as high computational costs, poor scalability, and insufficient accuracy across diverse material systems.

In this context, universal machine learning interatomic potentials (uMLIPs)[16-19] have emerged as a promising solution to bridge the gap between quantum mechanical methods and classical simulations. By leveraging large-scale datasets and advanced neural network architectures, uMLIPs achieve near-DFT-level prediction accuracy while maintaining the computational efficiency required for high-throughput research. Despite their promising potential, uMLIPs still exhibit limitations. For instance, some models struggle to extrapolate beyond the chemical space of their training data and fail to ensure physical consistency when calculating derived quantities such as forces or stresses[20-24]. Additionally, the predictive accuracy of these models varies across different material classes, highlighting the importance of carefully validating and selecting appropriate models for specific applications. To address these challenges, this study develops a systematic benchmark framework for SSEs to evaluate the performance of state-of-the-art uMLIPs—MatterSim[23], MACE[25], SevenNet[26], CHGNet[27], M3GNet[28], and ORBFF[29]—in complex material systems. Specifically, we assess their capabilities in three key areas:
1. Energy and force calculation: Accurate representation of potential energy surfaces and interatomic forces is fundamental to atomic simulations. Here, we compare the mean absolute error (MAE) and root mean square error (RMSE) of calculated energies and forces against DFT benchmarks, covering a wide range of halide, oxide, and sulfide electrolyte structures. Our analysis includes near-equilibrium and non-equilibrium structures sampled across extensive temperature and pressure ranges, ensuring comprehensive coverage of operational conditions.
2. Material property calculations: Beyond basic energy and force calculation, we delve into more complex material properties, including bulk modulus, shear modulus, formation energy, and energy above the convex hull (E_above_hull). These metrics provide valuable insights into the mechanical integrity, thermodynamic stability, and synthetic feasibility of candidate SSEs. By correlating uMLIPs calculation with DFT results, we aim to uncover patterns that reveal the strengths and weaknesses of models in capturing subtle yet crucial aspects of material behavior.
3. Lithium-ion diffusion simulations: Most critically, we investigate the ability of each uMLIP to reproduce lithium-ion diffusion coefficients across multiple SSE families—a direct measure of ionic conductivity. Using molecular dynamics (MD) simulations guided by uMLIPs, we compute mean square displacements (MSD) at various temperatures and extract diffusion constants. Comparisons with widely accepted DeepMD[30] benchmarks serve as a rigorous test of model fidelity in reproducing dynamic phenomena. Through these evaluations, we not only establish a clear hierarchy among the tested uMLIP models to determine their suitability for SSE research but also highlight the significance of the benchmark framework itself. As uMLIP models continue to evolve rapidly, systematically assessing their performance has become a critical challenge. Our framework provides a standardized tool for evaluating current models and lays the groundwork for the development and optimization of future uMLIPs.

The results demonstrate that MatterSim excels in nearly all evaluation metrics, particularly in calculating the thermodynamic, kinetic, and ion transport properties of SSEs with high precision and robustness. Even under challenging non-equilibrium conditions, MatterSim achieves exceptionally low MAEs for energy and force calculation, closely aligning with DFT and DeepMD benchmarks. It performs notably well in simulating bulk modulus, shear modulus, formation energy, energy stability, and lithium-ion diffusion pathways and activation barriers. Further analysis using MatterSim reveals the critical roles of S/Cl anion disorder levels and Na/Li arrangements in lithium-ion diffusion in $Li_6PS_5Cl$ and $Li_3YCl_6$. For instance, an S/Cl disorder level of approximately 40%-50% in $Li_6PS_5Cl$ optimizes diffusion pathway connectivity, significantly enhancing conductivity. In $Na_xLi_{3-x}YCl_6$, increasing lithium content expands diffusion channels and introduces local disorder, lowering migration energy barriers and improving ion transport performance. In contrast, other models such as SevenNet, MACE, CHGNet, and M3GNet exhibit notable limitations in various properties, while ORBFF suffers from systemic biases and poor generalization, resulting in lower reliability. These findings underscore the superiority of MatterSim in complex material systems, providing crucial theoretical support for designing high-performance SSEs through structural optimization. They also highlight the critical roles of training data quality and physical consistency in determining model performance.

In summary, this study provides a comprehensive evaluation of representative uMLIP models, revealing their respective strengths and limitations in SSE research. Through detailed analysis, we confirm the prioritized status of MatterSim for high-precision simulations, applicable to complex tasks encompassing energy landscapes, material properties, and ion transport mechanisms. More importantly, the SSE benchmark framework developed in this study offers a reliable and standardized method for systematically evaluating uMLIP models, effectively addressing gaps in model validation and comparison within the field. This achievement not only aids in optimizing current models but also provides a vital reference for the development of future uMLIP models. Ultimately, integrating advanced uMLIPs into routine computational workflows promises to revolutionize how we explore and design new materials, bringing us closer to realizing the full potential of SSEs in next-generation energy storage solutions.

## 2. Methods

### 2.1. uMLIPs description

In the context of studying alkali metal superionic conductors and evaluating uMLIPs as material calculators, gaining a deep understanding of various uMLIPs models is crucial. M3GNet, as a pioneering model in this field, employs a graph neural network architecture that integrates three-body interactions, capturing atomic interactions precisely through iterative updates of atom and bond features. Its training data encompasses a wide range of chemical elements and crystal structures, enabling it to perform well in structure optimization and static energy calculation, although it exhibits deviations in phonon property calculation[31]. CHGNet, based on a deep graph neural network with a sophisticated and efficient architecture, utilizes atomic magnetic moments to capture magnetic properties. It demonstrates strong convergence in structure optimization and moderate performance in energy calculation but shows significant errors in phonon property calculation[31]. MACE achieves high efficiency and accuracy through higher-order equivariant message-passing neural networks and E(3)-equivariant representations, leveraging atomic cluster expansions to enhance computational efficiency[32-34]. It excels in thermodynamic property calculations but shows larger errors in calculating elastic modulus-related properties. SevenNet, optimized for parallel computing based on NequIP[34], performs well in optimizing geometric structures and calculating thermodynamic properties but exhibits considerable errors in elastic modulus calculation. MatterSim combines the architectures of M3GNet and Graphormer to process material graphs, integrating extensive data through active learning and a unique data explorer, demonstrating the best overall performance across various property calculation. ORBFF directly outputs forces to improve computational efficiency, showing excellent performance in structure optimization and static energy calculation but exhibiting significant systematic biases in many thermodynamic property calculation. Benchmark testing of these advanced uMLIPs models reveals their strengths and limitations in different application scenarios. These results indicate that while all models achieve a certain level of accuracy in structure optimization, energy, and force calculation, the choice of model and the physical consistency of its computational methods significantly impact the accuracy of material property predictions.

### 2.2. Calculation of material properties based on uMLIPs

The process for calculating energy and forces based on uMLIPs is as follows: First, the process utilizes the RelaxCalc module within the Atomic Simulation Environment (ASE)[35] framework to perform structure optimization. During the initialization stage, users need to specify uMLIPs as the calculator and set key parameters such as the optimization algorithm (e.g., Fast Inertial Relaxation Engine-FIRE)[36], maximum optimization steps, convergence threshold (fmax), and whether to relax atomic positions or cell shape. Subsequently, the system converts the input crystal structure into an ASE atom object and loads the uMLIPs calculator to predict interatomic interaction potentials and their gradients (i.e., forces). During optimization, atomic positions and cell parameters are iteratively adjusted to minimize the total system energy while recording trajectory information at each step for subsequent analysis. The final output includes optimized structural parameters (e.g., lattice constants, volume, and angles), total system energy, atomic force distribution, and stress tensors. This process not only ensures physical consistency in energy and force calculations but also effectively reduces finite-size effects and non-physical distortions by incorporating tools such as trajectory observers and cell filters, providing a reliable foundation for studying the dynamic behavior of complex material systems like SSEs.

The procedure for calculating thermodynamic and mechanical properties of materials (including bulk modulus, shear modulus, formation energy, and convex hull energy) using uMLIPs is as follows: First, the material structure and correction energies are obtained from the Materials Project (MP)[37], and its CIF file is loaded to create a structure object. A universal atomic potential function is used as the ASE calculator, with SCF energy convergence and force convergence thresholds set to $10^{-5}$ eV and 0.05 eV/Å, respectively. The structure undergoes relaxation calculations to obtain the optimized structure and energy. Based on the relaxed structure and energy, a ComputedStructureEntry object is created by combining correction energies, and all entries within the chemical system (including MP data, derived data, and MatterSim results) are queried via the pymatgen interface. To ensure compatibility, MatterSim applies Hubbard U corrections for elements with localized d-electrons, aligning with MP data. Subsequently, a phase diagram is constructed to calculate formation energy and convex hull energy. Additionally, bulk and shear moduli are calculated by applying strain and analyzing stress responses. The final results are stored as JSON files in dictionary format.

The specific steps for calculating elastic moduli are as follows: During initialization, parameters such as the calculator, normal strains, shear strains, maximum force, whether to relax the structure, whether to consider equilibrium stress, and additional relaxation parameters are set. If structure relaxation is enabled, the crystal structure is first optimized to reach the lowest energy state. Then, a series of deformed structures are generated, and the stress response of each structure is calculated. By linearly fitting stress data under different strain states, the components of the elastic tensor are obtained. During fitting, equilibrium stress can optionally be considered, and the sum of squared residuals is calculated to evaluate accuracy. The final results are returned in dictionary format, containing key information such as the elastic tensor, Voigt-Reuss-Hill averaged shear modulus, bulk modulus, Young's modulus, and fitting residuals. This method not only reveals the elastic behavior of materials but also provides an efficient and precise tool for materials science research.

This study systematically conducted cross-system MD simulations of lithium-ion diffusion behavior in SSEs based on uMLIPs and Deep Potential (DP) methods. Using a self-developed MolecularDynamics module and the DeePMD framework, a standardized simulation workflow was established, covering 18 typical electrolyte systems across oxides, sulfides, and halides. Simulations were performed using 3×3×3 supercells (with more than 1200 atoms) under NVT/NPT ensembles, with temperature gradients ranging from 300 to 1100 K and ambient pressure conditions. The time step was set to 2.0 fs, and the evolution duration ranged from 1 to 10 ns. The NVT ensemble employed the Berendsen temperature coupling algorithm (τ_T=200 fs), while the NPT ensemble used a non-uniform Berendsen barostat algorithm (τ_P=2000 fs) to achieve pressure equilibrium. Atomic trajectories and thermodynamic parameters were recorded during the simulation, and the influence of local ordered states was eliminated through multi-frame averaging. The computational workflow consists of four key stages: (1) Initializing the atomic system and loading the potential function; (2) Configuring thermodynamic system parameters and couplers; (3) Performing dynamic iterations (e.g., 1000 steps corresponding to 2 ps); (4) Outputting physical quantity data and trajectory files. This framework supports both single-configuration and high-throughput multi-configuration modes, effectively linking structural evolution and transport properties of materials under equilibrium/non-equilibrium conditions.

The initial training dataset was constructed using the DP-GEN workflow[38], with independent training and exploration conducted for each SSE system. The specific process is as follows: Starting from randomly perturbed structures, short-time AIMD simulations[39] (50 ps, time step 2 fs) were performed at 300, 600, 900, and 1200 K, with snapshots output every 200 fs, generating 2500 initial structures for each system. Subsequently, active learning strategies were employed to increase data diversity, including the following three steps:

1. Training: The DP model was trained using DeePMD-kit, with embedding network and fitting network parameters set to [25, 50, 100] and [60, 60, 60], respectively, and a descriptor cutoff radius of 8 Å;
2. Exploration: Integrated DP models were used to run NPT ensemble MD simulations, selecting new structures with maximum force deviations *σf_max* falling between *σlow* (0.12 eV/Å) and *σhigh* (0.25 eV/Å);
3. Labeling: Single-point energy calculations were performed on the explored structures, and the results were added to the dataset.

After multiple iterations, each SSE system generated a final dataset of 3326 structures (80% for training, 20% for validation). In total, the dataset covered 18 SSE systems, involving 17 elements, and comprised 59,868 snapshots, ensuring broad coverage of chemical and structureal spaces. All data were labeled with energy, forces, and stresses calculated via DFT, providing a high-quality foundation for model training.

Simulations selected appropriate supercell sizes to minimize finite-size effects and were conducted under NVT or NPT ensembles for long durations (1-10 ns). Lithium-ion position changes were recorded, and MSD was calculated to analyze diffusion behavior:

$$\mathrm{MSD}(t) = \frac{1}{N}\sum_{i=1}^{N} |r_i(t) - r_i(0)|^2$$

The diffusion coefficient *D* was calculated from the slope of the MSD curve using the Einstein relation:

$$\mathrm{D} = \lim_{t\to\infty} \frac{\mathrm{MSD}(t)}{6t}$$

and linked to ionic conductivity via the Nernst-Einstein equation:

$$\sigma = \frac{Vz^2e^2}{k_BT} D$$

where $z$ is the charge number of lithium ions, $e$ is the elementary charge, $V$ is the simulation cell volume, $k_B$ is the Boltzmann constant, and $T$ is the temperature.

By comparing the results of the six uMLIPs with those of DeepMD, the reliability of the models was validated. Extending simulation durations (1-10 ns) and increasing supercell sizes significantly reduced finite-size effects, successfully capturing long-range lithium-ion diffusion behavior (e.g., continuous path evolution in the LGPS system). This multiscale simulation strategy not only significantly reduced computational costs but also provided a methodological basis for evaluating the applicability of uMLIPs in complex SSE systems. It offers crucial support for understanding lithium-ion diffusion mechanisms and designing high-performance SSE.

## 3. Results

### 3.1. Benchmark framework

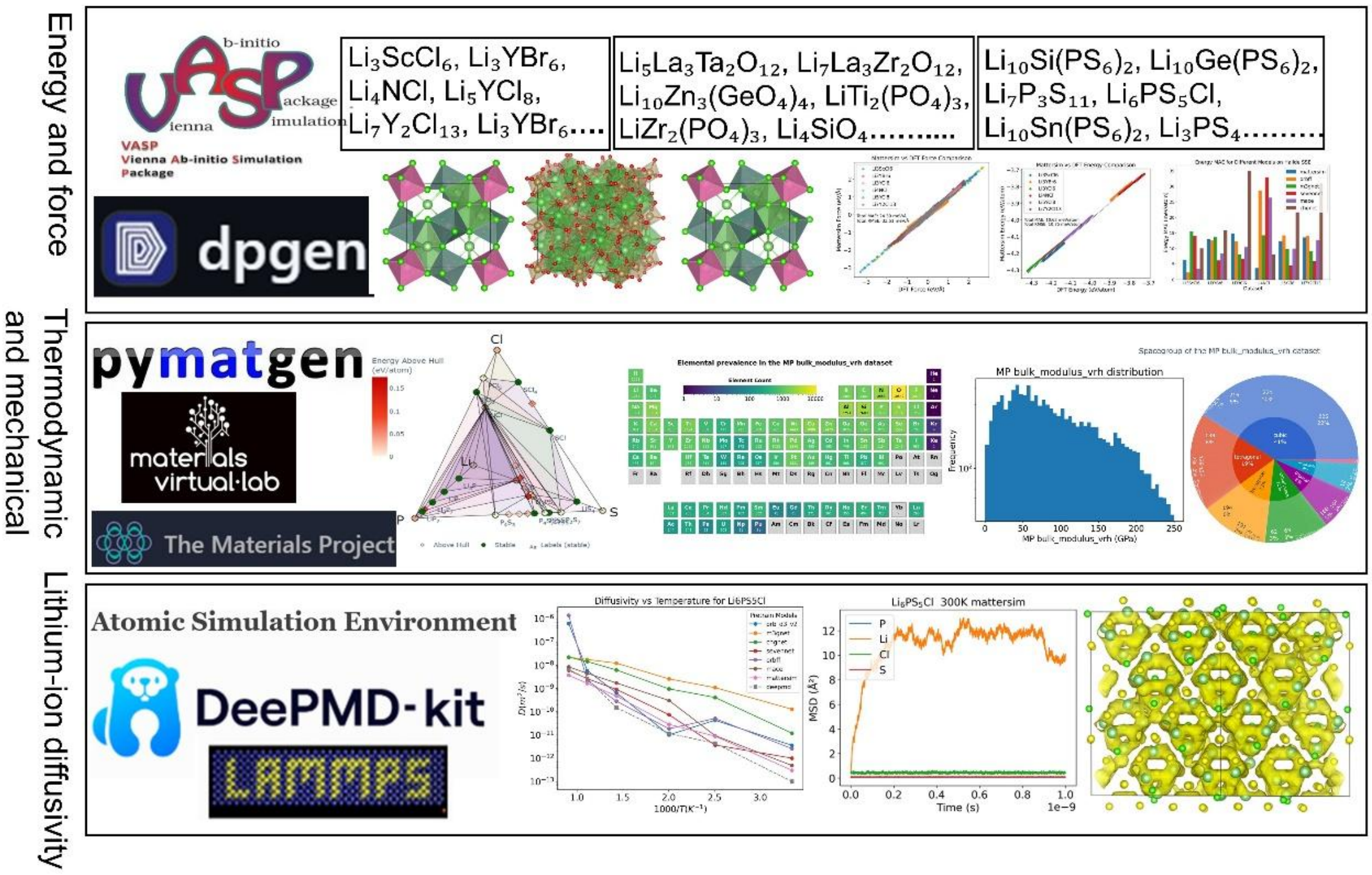

Figure 1 illustrates the benchmark framework used to evaluate uMLIPs in the calculation of material properties for SSEs, divided into three parts: 1. Energy and force calculation dataset: Generated through ab initio molecular dynamics (AIMD) and DeepGen, this dataset includes 1,980 snapshots covering sulfides, halides, and oxides across 18 systems composed of 17 elements. The data is labeled using DFT. 2. Thermodynamic and mechanical properties dataset: Constructed based on the MP database using the MP + Pymatgen + DIRECT sampling algorithm, this dataset include formation energy, bulk modulus, shear modulus, and E_above_hull. 3. Lithium-ion diffusivity dataset: Obtained through long-time MD simulations performed using ASE + DeepMD + LAMMPS, providing data on lithium-ion diffusivity.

This paper develops a specialized benchmark framework for evaluating the performance of uMLIPs in calculating properties of SSEs. As shown in Figure 1, the framework consists of two core components: (1) a dataset for energy and force calculation across 18 typical SSE systems, covering sulfides, halides, and oxides; and (2) a dataset describing key material properties of SSEs, including bulk modulus, shear modulus, E_above_hull, formation energy, and lithium-ion diffusivity. When screening SSE materials, several critical properties must be considered. Stability ensures the practical applicability of the material; lithium-ion diffusivity directly determines the charge-discharge performance of batteries; electronic insulation prevents internal short circuits; electrochemical stability ensures the material remains stable within the operating voltage range of the battery; mechanical properties facilitate good contact between electrodes and materials; and density is closely related to the energy density of the battery. However, since density data is relatively easy to obtain, electronic insulation cannot be directly calculated using uMLIPs models, and electrochemical stability essentially falls under energy and formation energy calculations, this study focuses on evaluating five properties—bulk modulus, shear modulus, E_above_hull, formation energy, and lithium-ion diffusivity—to comprehensively assess the performance of uMLIPs in SSE research.

To comprehensively evaluate the accuracy of uMLIPs in calculating energy and forces in SSE systems, this study constructs a dataset encompassing both equilibrium and non-equilibrium structures. The dataset is generated using AIMD and DeepGen methods, ensuring its physical consistency and chemical diversity. The data generation process involves two stages: first, initial structures are randomly perturbed and sampled through short-range AIMD simulations; then, NPT ensemble simulations are conducted over a temperature range of 100–1100 K and a pressure range of 0–2 GPa to capture non-equilibrium and near-equilibrium structures. The final training set includes 1,980 snapshots, covering three categories of typical SSEs (six systems per category): sulfides, halides, and oxides, involving 18 systems composed of 17 elements. All structures are annotated for energy and forces using VASP software, employing PAW pseudopotentials[40] with a plane-wave cutoff energy of 600 eV. Atomic coordinate relaxation convergence is set to 0.01 eV/Å, and standardized Brillouin zone sampling is performed using KSPACING=0.3. The wide temperature-range excitation strategy significantly enhances the coverage of configuration space, while the unified quantum chemistry calculation protocol ensures high precision and physical consistency of energy, force, and stress labels.

To further validate the performance of uMLIPs in calculating thermodynamic and mechanical properties, this study constructs a three-tiered dataset based on formation energy, bulk modulus, shear modulus, and E_above_hull data from the MP database. For formation energy and E_above_hull data, 77,680 structures (50% of the total) are sampled from 155,361 structures using the hierarchical clustering algorithm DIRECT[41], serving as a general testing dataset that covers the full element space and crystal structure diversity to verify the global generalization ability of the models. Subsequently, lithium-containing compounds (10,367 structures for formation energy and E_above_hull) and typical SSE structures (606 structures for formation energy and E_above_hull) are screened to construct lithium-containing compound testing datasets and lithium-containing SSE testing datasets, focusing on lithium-based systems and key areas of energy materials for specialized performance validation. For bulk and shear moduli, three levels of testing datasets are constructed based on 12,334 original structures: the general testing dataset includes 10,812 non-sulfide/halide/oxide structures while retaining 50% lithium-containing compounds to maintain elemental diversity; the lithium-containing compound testing dataset extracts 422 lithium-containing structures; and the lithium-containing SSE testing dataset further focuses on 153 typical sulfide/halide/oxide electrolytes and their decomposition products. This hierarchical testing framework systematically refines evaluation scenarios, validating the applicability of uMLIPs in calculating thermodynamic and mechanical properties and providing an important basis for cross-material system model optimization.

In SSE research, the accurate calculation of lithium-ion diffusivity is crucial because it directly affects the charge-discharge efficiency and power density of batteries. High diffusivity means lithium ions can move quickly within the electrolyte, enabling efficient charge-discharge processes, which is significant for improving battery response speed and meeting high-power application demands. By accurately calculating lithium-ion diffusivity, researchers can identify materials with excellent ionic conductivity, providing theoretical guidance for designing high-performance solid-state batteries. However, traditional computational methods have significant limitations: empirical force field-based simulations are fast but lack accuracy, struggling to maintain consistent precision across different material structures and temperature ranges; AIMD methods, while highly accurate, are computationally expensive, typically limited to systems of a few hundred atoms and tens of femtoseconds. Therefore, evaluating the performance of uMLIPs models in calculating lithium-ion diffusivity in SSEs is particularly important. To this end, this study systematically evaluates advanced uMLIPs models in 18 typical SSE systems, including oxide, sulfide, and halide electrolytes. We use these models to calculate lithium-ion diffusivity at various temperatures (300 K, 400 K, 500 K, 700 K, 900 K, and 1100 K), with DeepMD results serving as the reference standard. To ensure computational accuracy, NPT and NVT ensembles are employed, and long-duration (1–10 nanoseconds) MD simulations are conducted under ambient pressure. Each SSE uses a 3×3×3 supercell containing over 1,200 atoms to minimize finite-size effects and more realistically reflect ion diffusion behavior in actual materials. Additionally, considering the influence of order/disorder states on lithium-ion diffusion, we average multi-frame data to enhance statistical reliability.

The SSE benchmark framework developed in this study is comprehensive, consistent, and diverse, providing a systematic tool for evaluating the performance of uMLIPs models in SSE property calculations. The framework encompasses multidimensional assessments, including structure optimization, energy and force calculations, mechanical properties, thermodynamic properties, and kinetic properties (e.g., lithium-ion diffusivity), ensuring a thorough examination of model performance. Consistency is achieved by adopting the PBE functional consistent with the uMLIPs model training data, effectively reducing errors caused by differences in exchange-correlation functionals. Diversity is reflected in the testing dataset, which covers a wide range of chemical elements and crystal structures, verifying the generalization ability of the models across different material systems. Furthermore, the framework not only quantifies model performance using statistical metrics such as MAE and RMSE but also visually presents prediction deviations through error distribution plots, enabling researchers to gain deeper insights into the strengths and limitations of the models. This design provides clear guidance for developing new uMLIPs models while revealing directions for improving existing ones, thereby advancing the application of machine learning in materials science.

### 3.2. Evaluation of uMLIPs on energy and forces

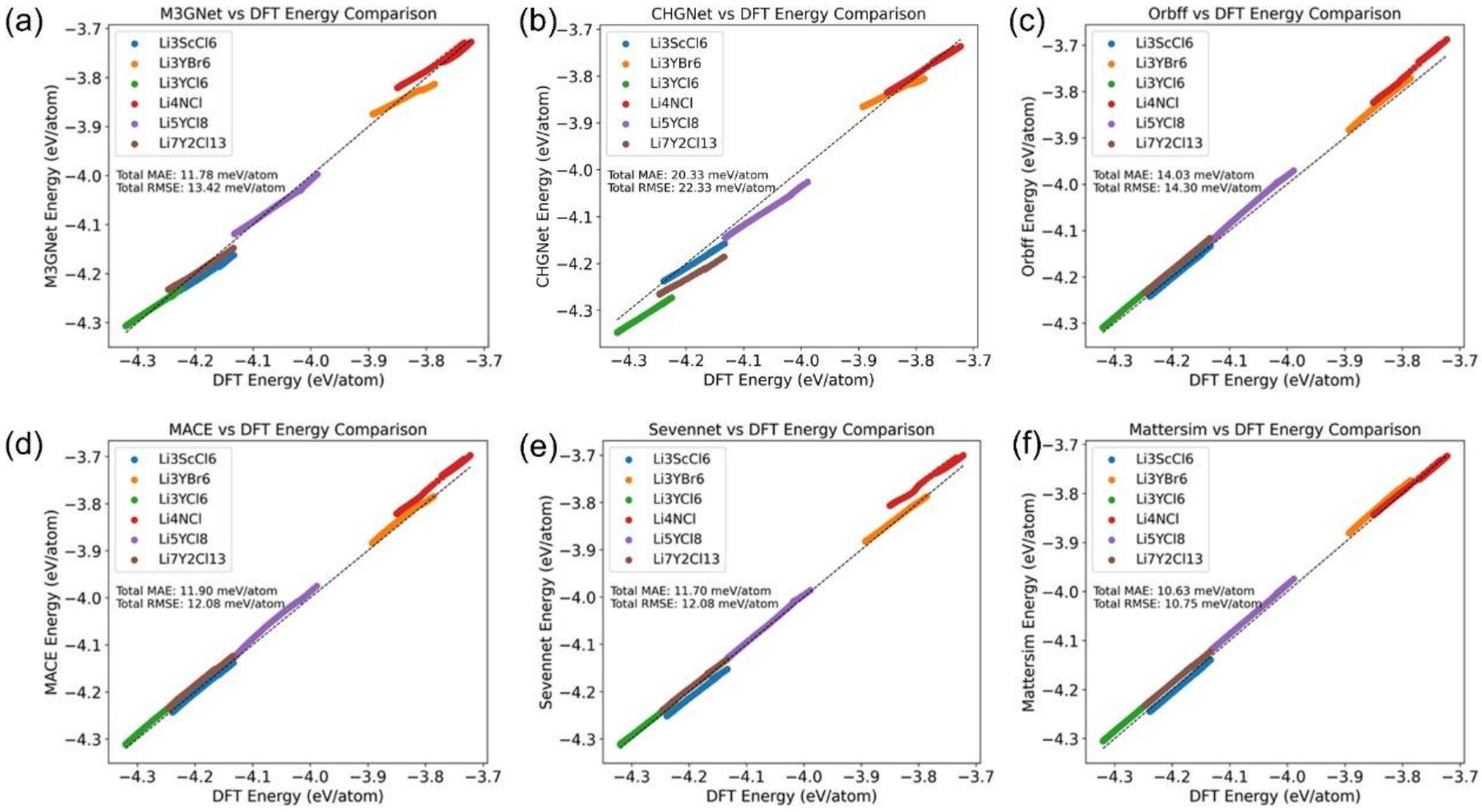

Figure 2 evaluates the performance of six uMLIP models in calculating the energy of halide electrolytes at different structures sampled from heated DeePMD trajectories ranging from 100 K to 1100 K, compared to DFT.

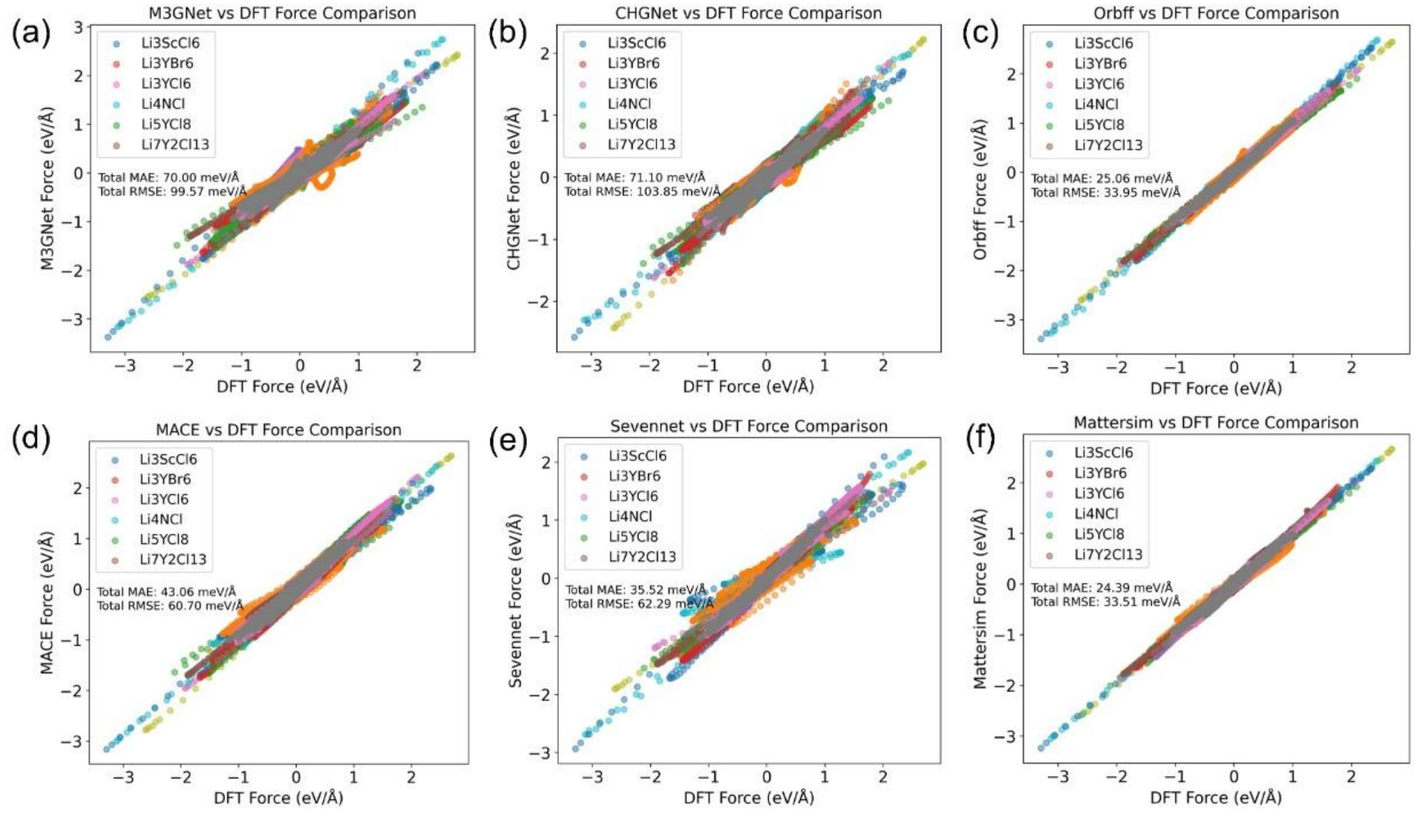

Figure 3 shows the comparison of errors between six uMLIP models and DFT calculations when computing forces on different structures of halide electrolytes. These structures were extracted from DeePMD trajectories that were heated from 100K to 1100K.

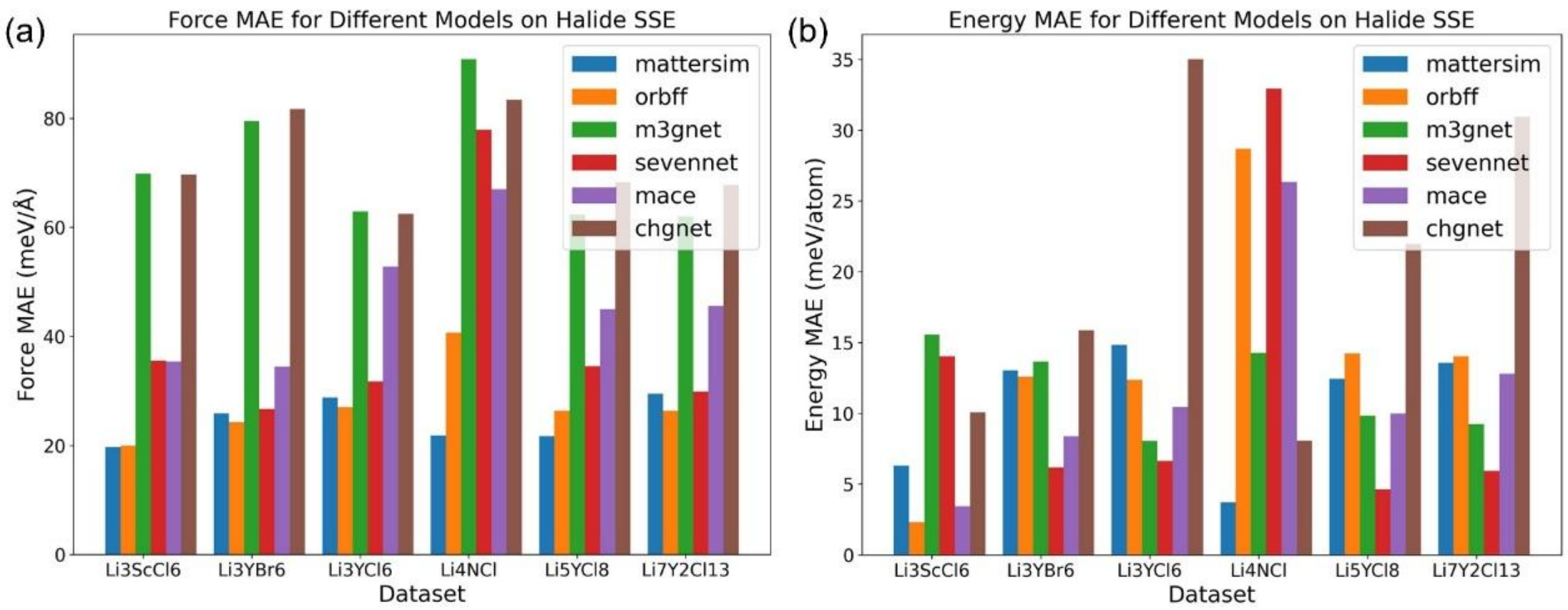

Figure 4, Evaluation of the performance of six uMLIPs models in calculating energy and forces on typical halide electrolytes using heated DeePMD trajectories. These trajectories were sampled across a temperature range from 100K to 1100K and include a large number of non-equilibrium structures.

Figures 2 to 4 correspond to the non-equilibrium energy and force tests for halides, showcasing the accuracy of the six uMLIP models in calculating energy and forces for different structures of halide electrolytes. Specifically, Figures 2 and 3 compare the calculated energies and forces with DFT results. In this test, MatterSim achieved the lowest MAE for energy and force calculation, converging to 10.63 meV/atom and 24.39 meV/Å, respectively. MACE's energy and force MAEs converged to 11.90 meV/atom and 43.06 meV/Å, respectively. SevenNet’s MAEs were 11.7 meV/atom and 35.52 meV/Å, respectively. ORBFF’s MAEs were 14.03 meV/atom and 25.06 meV/Å, respectively. CHGNet’s MAEs were 20.33 meV/atom and 71.01 meV/Å, respectively. M3GNet’s MAEs were 11.78 meV/atom and 70.00 meV/Å, respectively. Supplementary Figures S1 to S6 correspond to the non-equilibrium energy and force tests for sulfides and oxides, showing that MatterSim and ORBFF perform the best, followed by SevenNet and MACE, with CHGNet and M3GNet performing the worst.

To further evaluate model performance, we also tested these uMLIPs in near-equilibrium positions across oxides, sulfides, and halides. Figures S7 to S9 show the energy calculation results relative to DFT, while Figures S10 to S12 show the force calculation results. Figures S13 and S15 compare the performance of these uMLIP models across 18 systems of oxides, sulfides, and halides. Energy tests reveal that all uMLIP models perform best on sulfides, with MAE values ranging from 6.6 to 9.8 meV/atom, and worst on oxides, with MAE values ranging from 18.43 to 35.53 meV/atom. Force tests show that all uMLIP models perform best on halides, with MAE values ranging from 14.54 to 24.07 meV/Å, and worst on oxides, with MAE values ranging from 52.80 to 70.67 meV/Å. MatterSim, SevenNet, and ORBFF perform the best, followed by MACE, with CHGNet and M3GNet performing the worst. All uMLIP models show better performance in near-equilibrium force calculation than in non-equilibrium. For energy, the improvement in near-equilibrium calculation is less pronounced compared to force calculation.

In summary, these tests demonstrate that MatterSim excels not only in near-equilibrium but also in non-equilibrium, which is critical for accurately simulating the dynamic behavior of SSEs, such as ionic conductivity. These results highlight MatterSim as a powerful tool for high-precision and efficient calculator of solid-state electrolyte dynamics, providing important support for the development and optimization of SSE.

### 3.3. Evaluation of uMLIPs on thermodynamic and mechanical properties

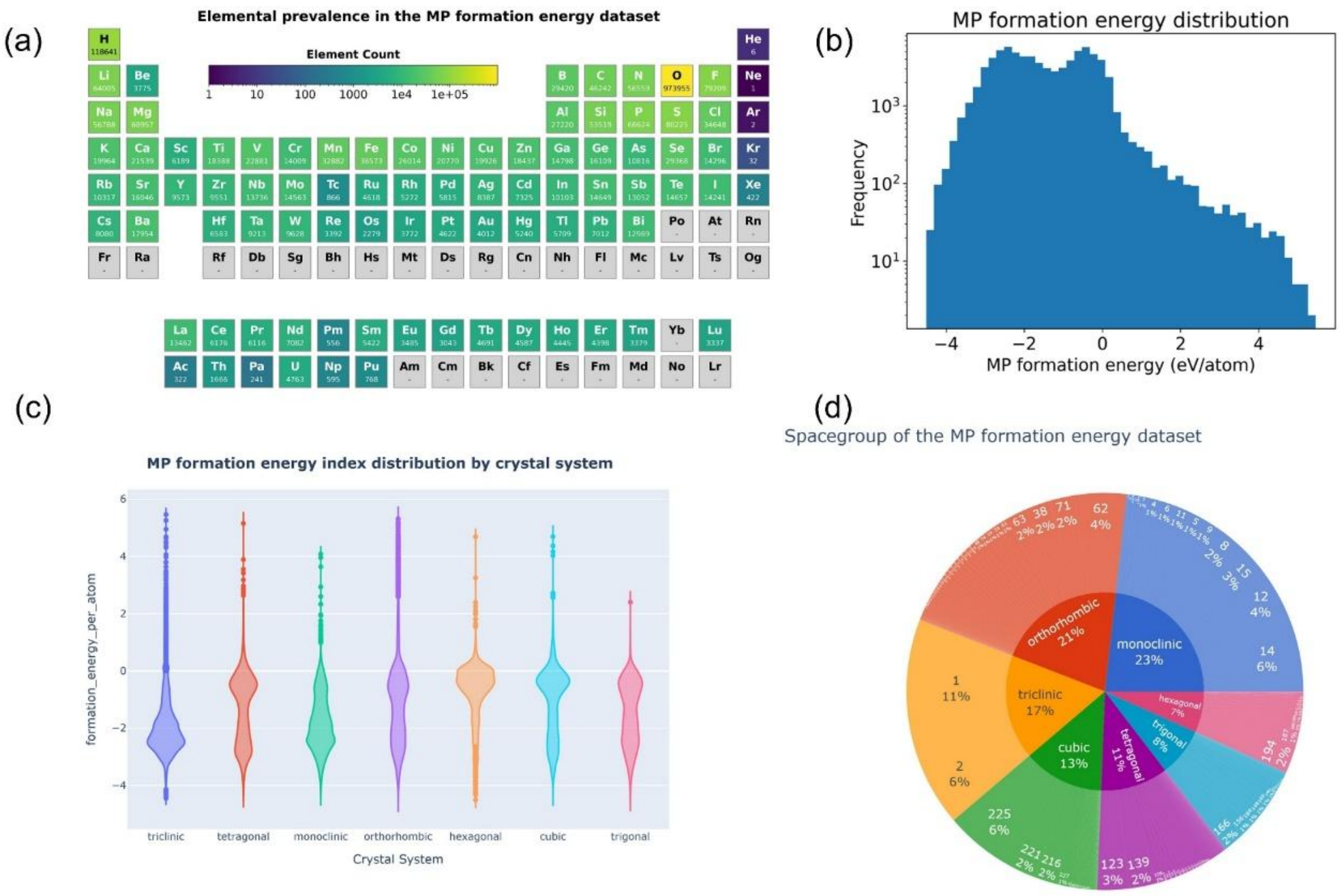

Figure 5 presents the visualization analysis results of 77,680 structures sampled from 155,361 structures in the MP database using the hierarchical clustering algorithm Direct, as a general dataset. The specific contents include:(a) Visualization of element distribution;(b) Frequency distribution histogram of formation energy values;(c) Distribution of formation energy values classified by crystal system;(d) Pie chart of the formation energy dataset classified by crystal system and corresponding space groups.

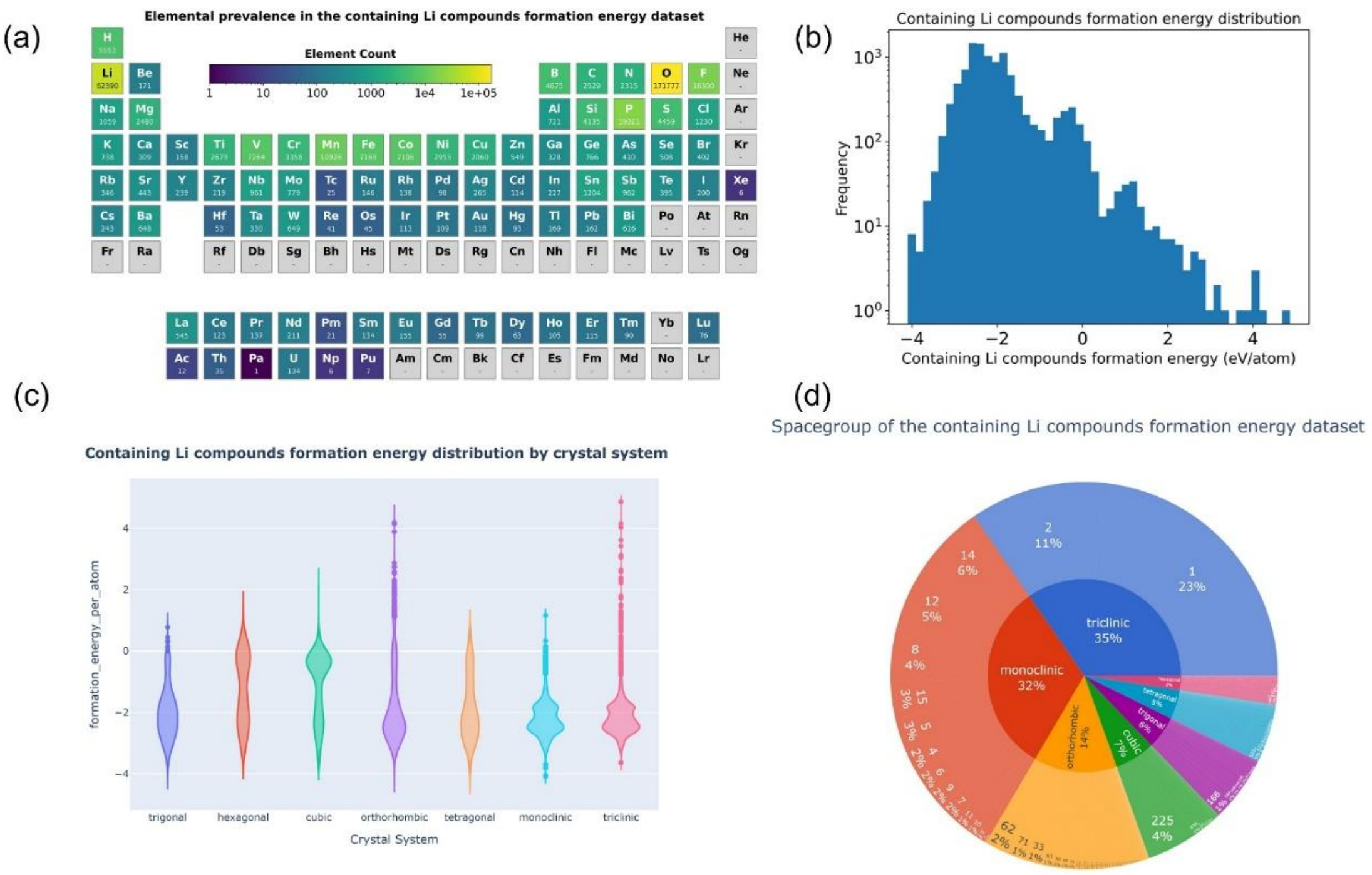

Figure 6 presents the visualization analysis results of all lithium-containing compound structures selected from the remaining 77,680 structures after sampling via the hierarchical clustering algorithm Direct from the 155,361 structures in the MP database. These structures form the lithium-containing compound dataset. The specific contents include: (a) Visualization of element distribution; (b) Frequency distribution histogram of formation energy values; (c) Distribution of formation energy values classified by crystal system; (d) Pie chart of the formation energy dataset classified by crystal system and corresponding space groups.

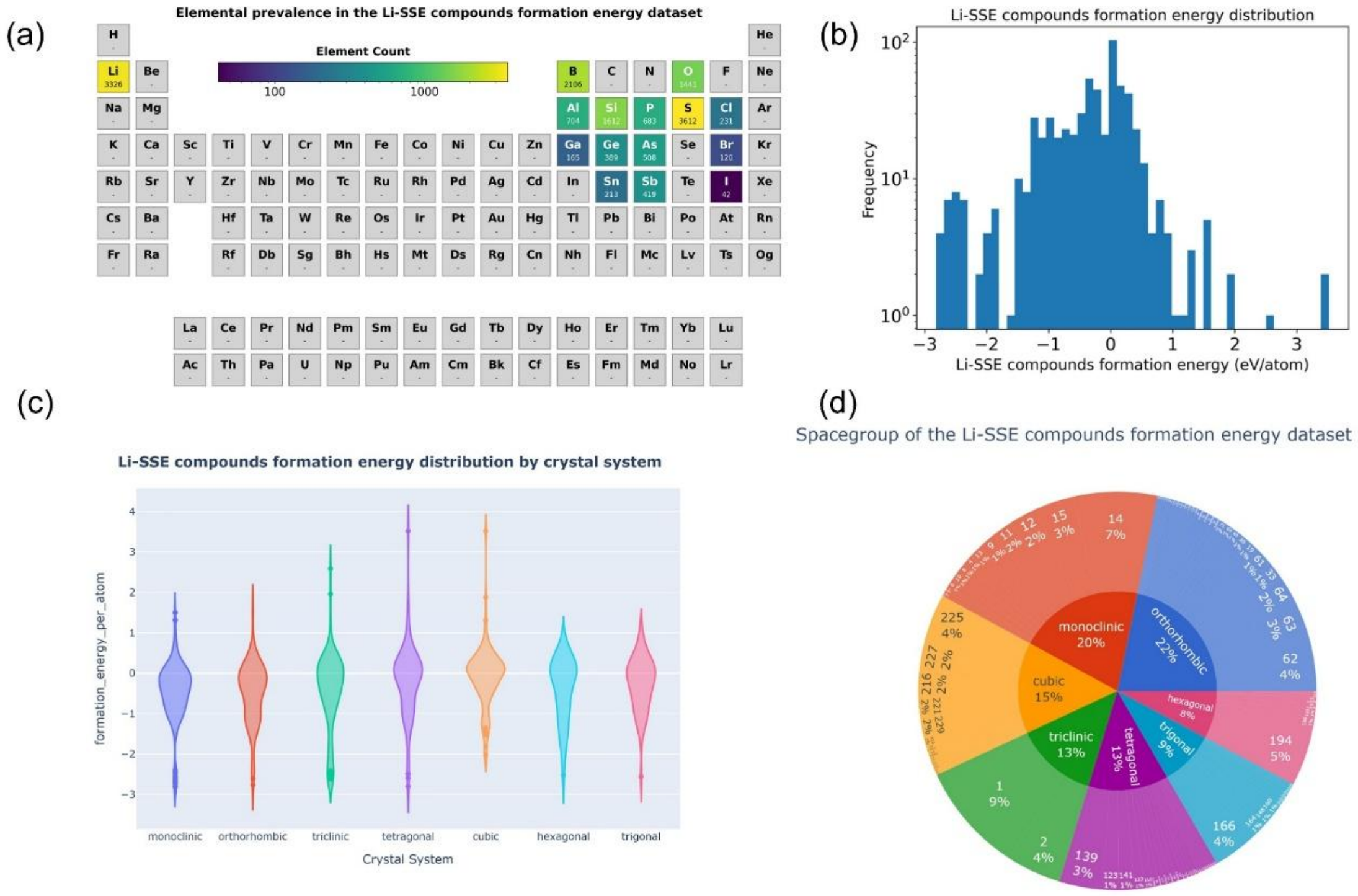

Figure 7 presents the visualization analysis results of all common lithium-containing SSE structures selected from the remaining half of the 155,361 structures in the MP database after sampling via the hierarchical clustering algorithm Direct. These structures form the lithium-containing SSE dataset. The specific contents include:(a) Visualization of element distribution;(b) Frequency distribution histogram of formation energy values;(c) Distribution of formation energy values classified by crystal system;(d) Pie chart of the formation energy dataset classified by crystal system and corresponding space groups.

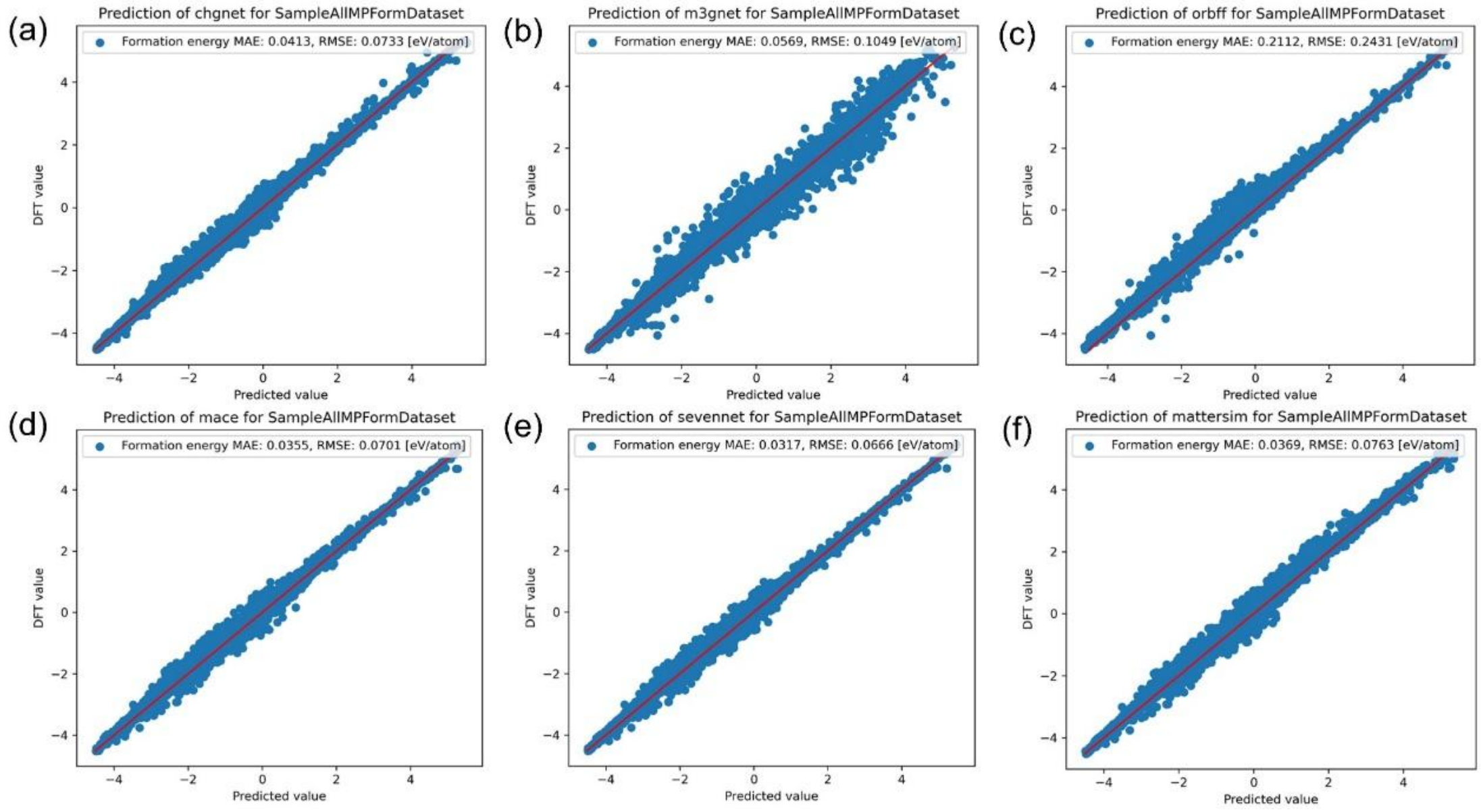

Figure 8 compares the computational results of six uMLIPs models with DFT results on the general dataset of formation energies.

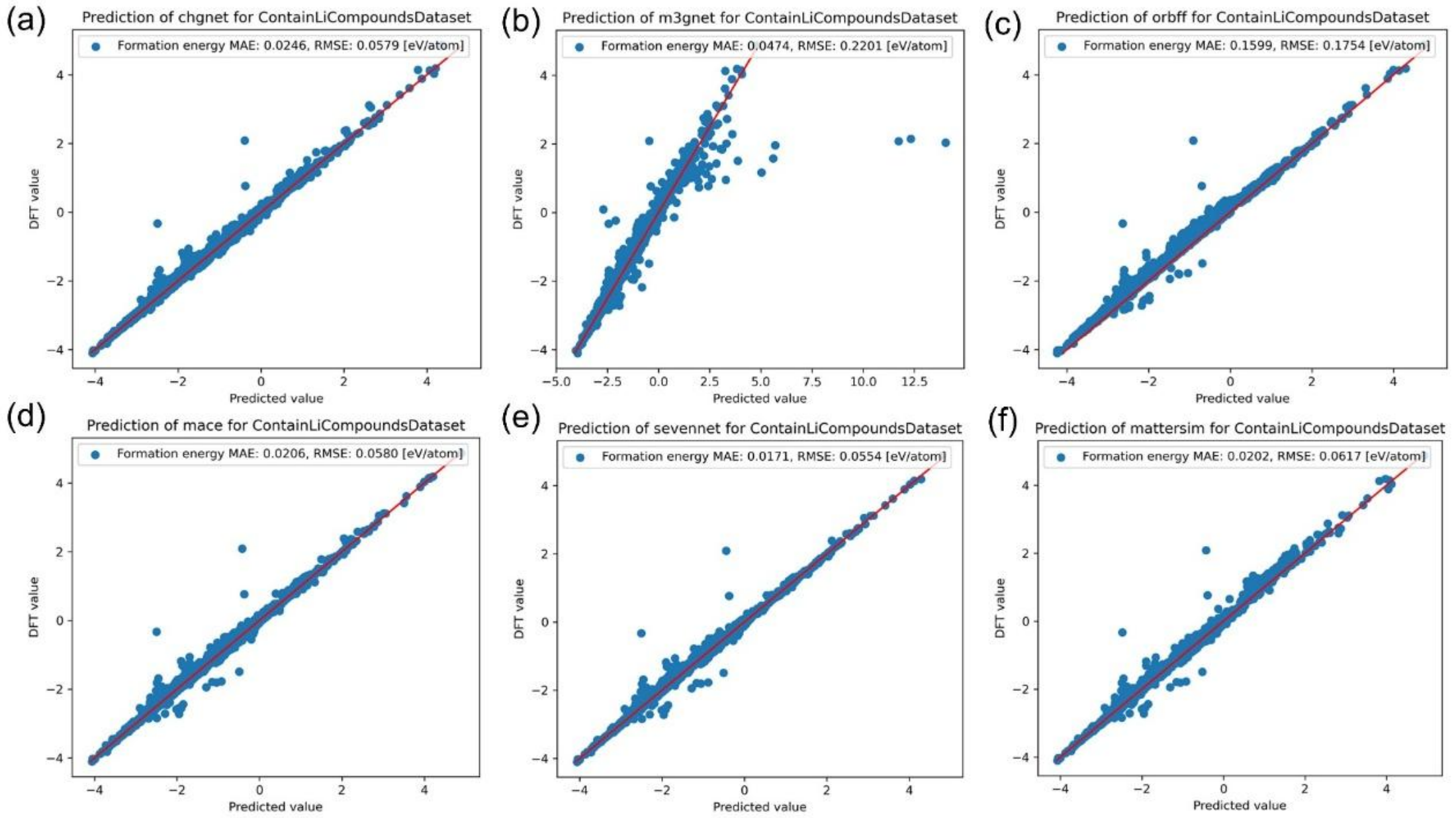

Figure 9 compares the computational results of six uMLIPs models with DFT results on the formation energy dataset of lithium-containing compounds from MP.

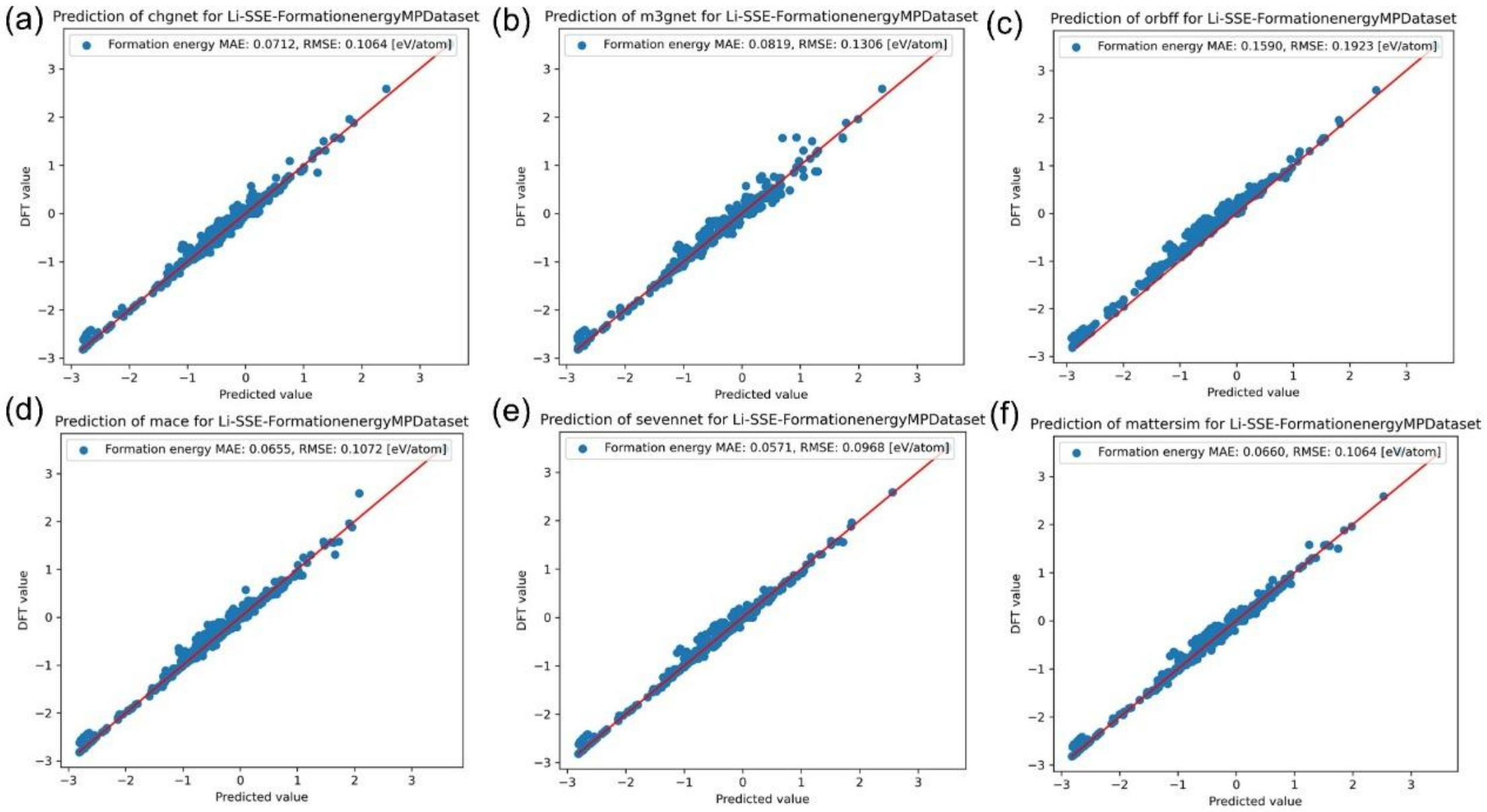

Figure 10 evaluates the calculated formation energies of six uMLIPs models on the MP lithium-containing SSE dataset, comparing them with DFT results.

Figures 5–7 and S28–30 reveal the structural characteristics of the hierarchical datasets for formation energy and E_above_hull through visualizations of element distribution, formation energy statistics, crystal systems, and space group classifications. In contrast, Figures 8–10 systematically compare the computational errors of uMLIPs models against DFT results across the general dataset (Figure 8), lithium-containing compound dataset (Figure 9), and lithium-containing SSE dataset (Figure 10). These comparisons illustrate the performance trends of the models as they transition from broad material spaces to specific functional systems. Additionally, for E_above_hull data, Supplementary Figures S31–S33 provide a systematic comparison of the error differences of uMLIPs models relative to DFT across the general dataset, lithium-containing compound dataset, and lithium-containing SSE dataset: Figure S31 highlights the global performance on the general dataset, Figure S32 emphasizes prediction deviations in lithium-containing systems, and Figure S33 focuses on the variation patterns of key material properties within SSE structures. Supplementary Figures S16–S18 (bulk modulus) and S22–S24 (shear modulus) validate the structural representativeness through element distribution, modulus statistics, and crystal system classifications. Meanwhile, Figures S19–S21 (bulk modulus) and S25–S27 (shear modulus) systematically compare the computational errors of uMLIPs models relative to DFT results across different datasets, revealing the performance trends of uMLIPs models as they shift from global generalization to specialized applications in energy materials. This hierarchical testing framework refines evaluation scenarios step by step, verifying that the computational accuracy of uMLIPs models significantly depends on property types (e.g., thermodynamic stability and mechanical properties). It provides crucial guidance for optimizing machine learning potentials across material properties and establishes a multiscale validation benchmark for assessing the reliability of machine learning potentials in complex material systems.

Through three hierarchical datasets (general dataset, lithium-containing compound dataset, and lithium-containing SSE dataset), we systematically evaluated the computational performance of uMLIPs on four key material properties: formation energy, bulk modulus, shear modulus, and E_above_hull. Overall, the MatterSim model demonstrated the highest comprehensive accuracy and stability, while ORBFF exhibited significant systematic deviations in all property calculations, with computed values generally lower than DFT results. Specific analyses are as follows: As shown in Figures 8–10, SevenNet performed best in the calculation of formation energy; MACE and MatterSim showed similar performance; CHGNet outperformed M3GNet, while M3GNet exhibited instability. In the general dataset, M3GNet suffered from high error dispersion (213 structures exceeding five times the average error) and showed significant deviations from DFT reference values in the lithium-containing compound dataset, leading to its lower ranking. ORBFF, due to directly outputting forces instead of energy derivatives, caused physical inconsistencies, resulting in the largest computational deviations. Further analysis of the differences between the second-worst-performing M3GNet and the best-performing SevenNet in the general dataset (as shown in Supplementary Figures S38–S41) revealed that 213 structures in M3GNet, 43 in SevenNet, 75 in ORBFF, 73 in MatterSim, 50 in MACE, and 124 in CHGNet had computational errors exceeding five times the average error. Specifically, the 213 structures in M3GNet covered almost all elements in the periodic table, except for some lanthanides and actinides, with calculated formation energy values ranging from -5 to 15 eV/atom, while actual DFT-calculated values ranged from -5 to 5 eV/atom, indicating that M3GNet overestimated the formation energies of many structures. In contrast, the best-performing SevenNet had only 43 structures with computational errors exceeding five times the average error, and these structures covered a narrower range of elements and crystal systems, with calculated values closer to DFT results.

By comparing Supplementary Figures S19–S21 and S25–S27, it is evident that MatterSim performed optimally and most stably in the calculation of bulk and shear moduli. MACE and SevenNet showed similar performance but were inferior to MatterSim across almost all three levels of datasets. CHGNet and M3GNet exhibited similar computational accuracy, but M3GNet had extreme outliers in the general dataset, such as errors in bulk modulus exceeding 50 GPa. ORBFF, due to insufficient generalization capability, systematically underestimated modulus values, particularly performing poorly in shear modulus calculations.

By comparing Supplementary Figures S31–S33, it is clear that SevenNet exhibited the lowest MAE and RMSE in E_above_hull calculations, demonstrating the best computational performance. Specifically, SevenNet had only 63 structures with computational errors exceeding five times the average error, indicating fewer large-error results. Further analysis using Supplementary Figures S34–S37 revealed that SevenNet had the smallest number of elements with computational errors exceeding five times the average error, and these errors were mainly concentrated in regions with high E_above_hull values, suggesting that SevenNet had certain inaccuracies in handling high E_above_hull value structures across different crystal systems. In contrast, MACE and MatterSim showed relatively similar performance in the global dataset, but differed in the number of structures with computational errors exceeding five times the average error, with 66 and 90 structures, respectively. Further observation of Figures S34–S37 showed that MACE performed better at lower E_above_hull values, while larger computational errors mainly occurred in structures corresponding to high E_above_hull values; MatterSim exhibited larger computational errors across a wider range of E_above_hull values. M3GNet, due to insufficient training data coverage, had the highest number of structures (229) with computational errors exceeding five times the average error, and these errors were widely distributed across most crystal systems, especially in cases where E_above_hull values were overestimated. ORBFF systematically failed in complex phase stability calculations, possibly due to its design of directly outputting forces rather than calculating them through energy derivatives. Although this approach improved efficiency, it may have led to a lack of physical consistency, thereby affecting the accuracy of energy calculations. Additionally, ORBFF's insufficient generalization capability on unseen data resulted in systematic deviations in calculations for new material systems or structures.

A deeper analysis of model performance differences using Supplementary Figures S34–S41 revealed that ORBFF's failure was primarily due to its non-physical design of directly outputting forces, which led to poor consistency in energy derivatives, specifically manifesting as systematic underestimation of formation energy and E_above_hull. M3GNet's discrete errors were related to insufficient training data coverage, with error-exceeding structures widely distributed across nearly all elements in the periodic table (except for some lanthanides and actinides) and all crystal systems. In contrast, SevenNet and MatterSim performed well because they imposed physical constraints (such as energy-force consistency) on potential energy surfaces and could represent rare elements and complex crystal systems more equitably, thereby minimizing the number of structures with excessive errors and achieving optimal predictive performance. In summary, this study revealed the performance ranking of uMLIPs in cross-property calculation as MatterSim > SevenNet ≈ MACE > CHGNet > M3GNet > ORBFF, and clarified the critical impact of model design (e.g., energy derivative consistency, training data coverage) on prediction accuracy and generalization capability.

### 3.4. Evaluation of lithium-ion diffusivity based on uMLIPs

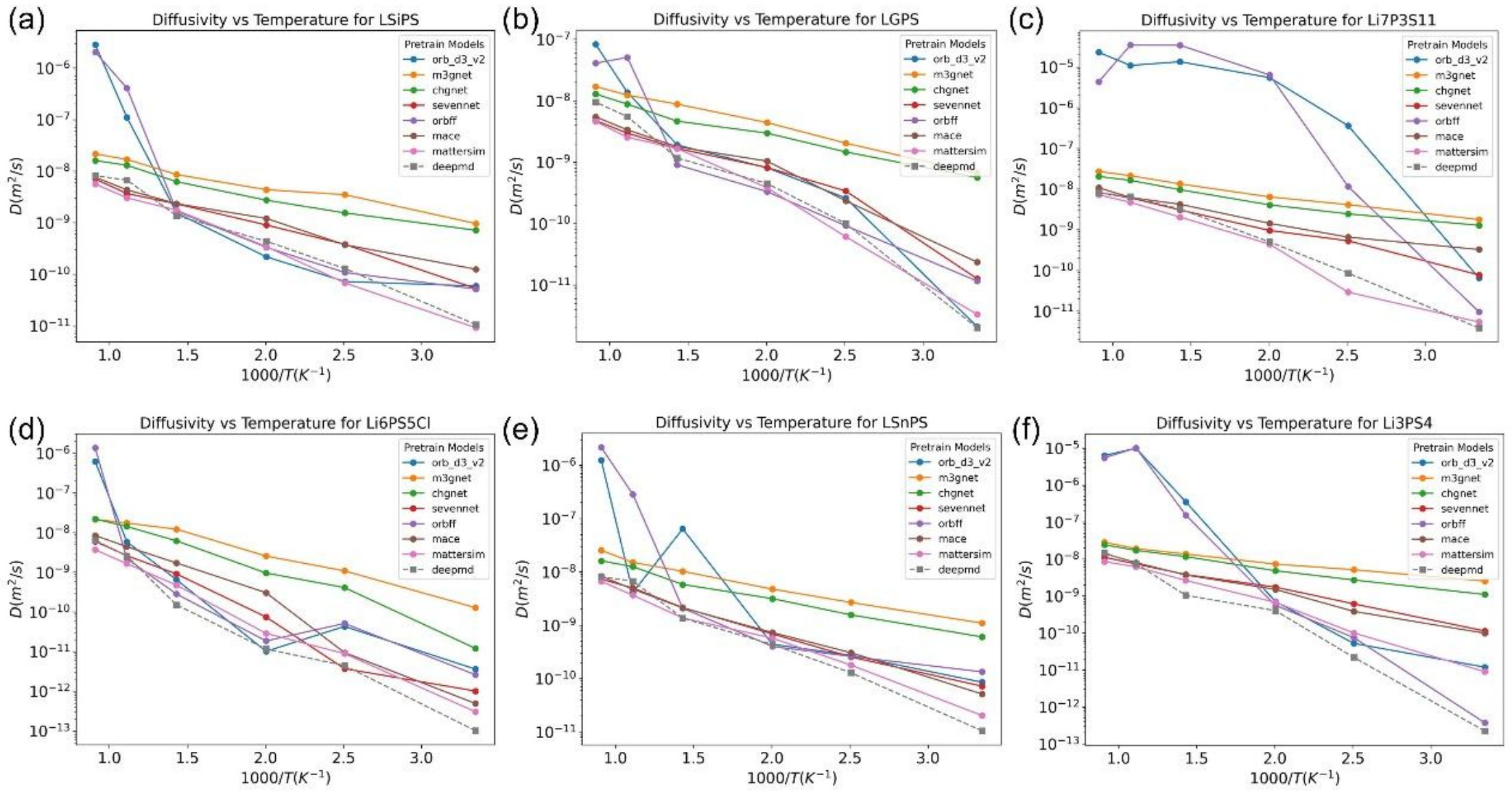

Figure 11 shows the comparison of lithium-ion diffusivity calculations between uMLIPs models and the DeepMD method for six typical sulfide SSEs.

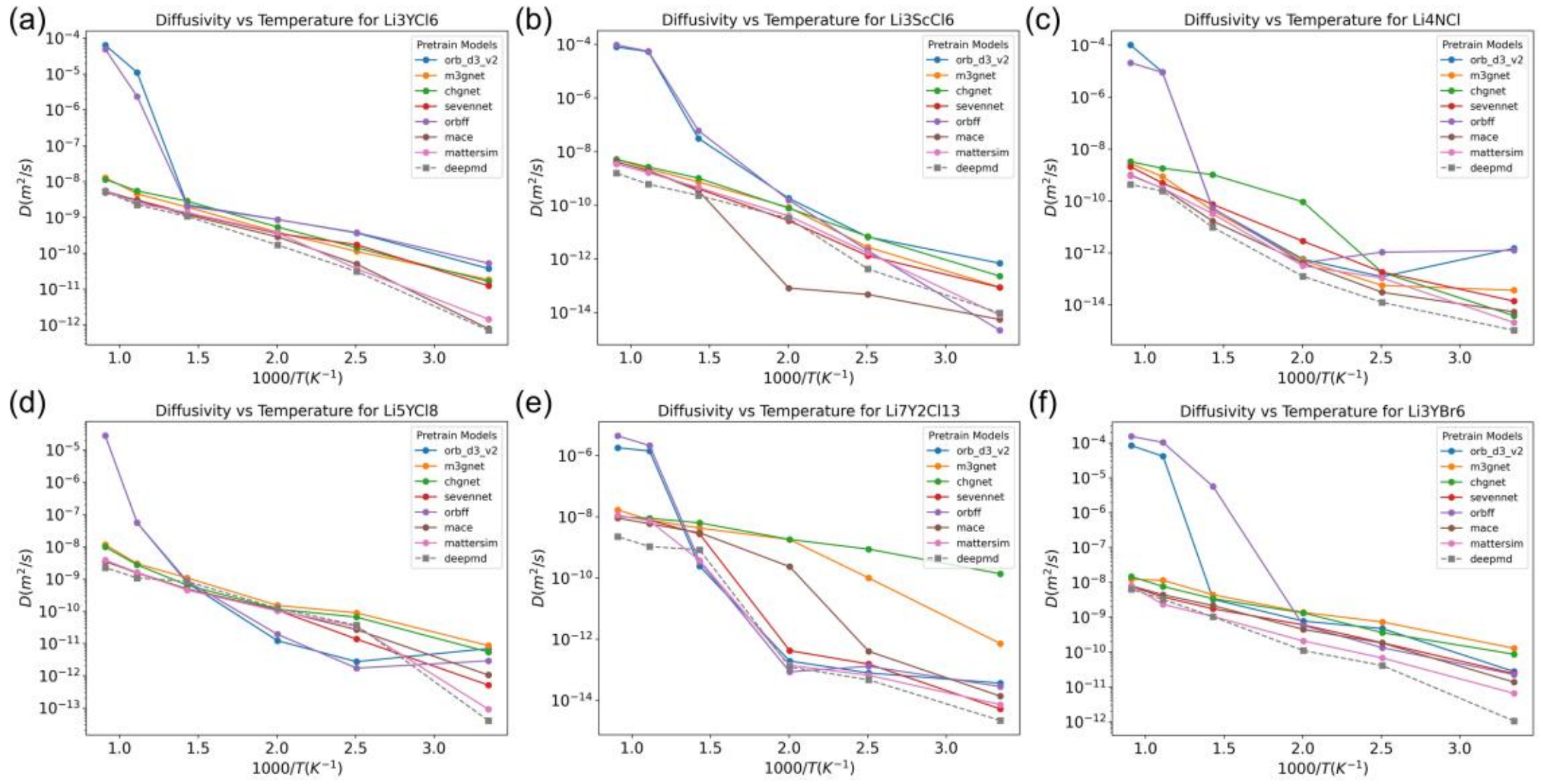

Figure 12 shows the comparison of lithium-ion diffusivity calculations between uMLIPs models and the DeepMD method for six typical halide SSEs.

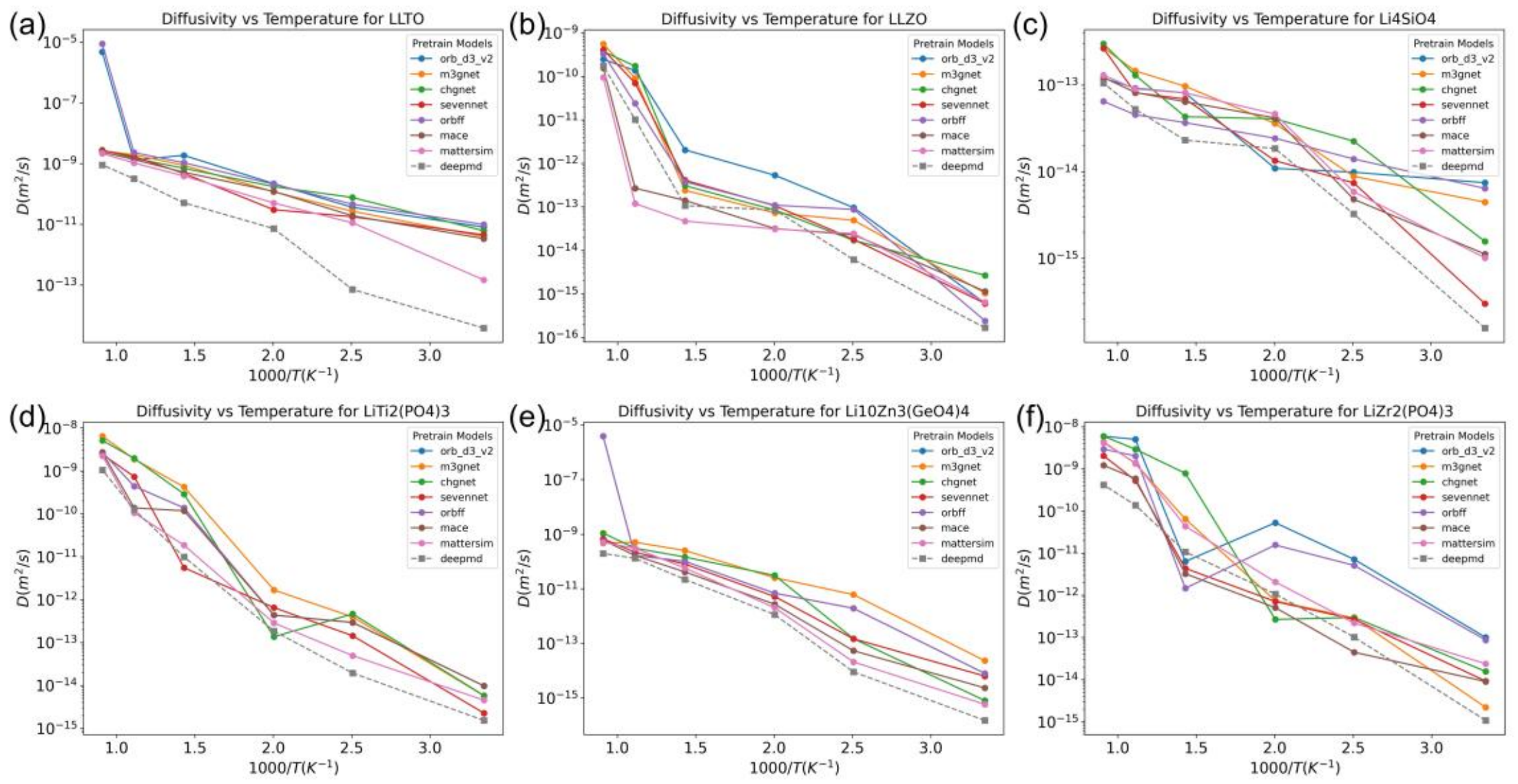

Figure 13 shows the comparison of lithium-ion diffusivity calculations between uMLIPs models and the DeepMD method for six typical oxide SSEs.

Figures 11 to 13 present the comparison of lithium-ion diffusivity calculated by uMLIPs models in typical sulfide, halide, and oxide SSEs against DeepMD reference values. The analysis shows that MatterSim performs exceptionally well across all temperatures, particularly in the low-temperature region, where its results closely match the DeepMD reference values, demonstrating high accuracy and reliability. MACE and SevenNet perform adequately in the mid-to-high temperature range but show reduced accuracy in the low-temperature region. CHGNet and M3GNet exhibit poor performance at all temperatures, with significant deviations from the reference values, especially in the low-temperature region. ORBFF's performance is inconsistent, with results severely deviating from the DeepMD reference values at high temperatures, while showing mixed results in the mid-to-low temperature range. To further validate these findings, Supplementary Figures S42 to S47 present the results of MD simulations conducted using MatterSim and SevenNet on typical sulfide, halide, and oxide SSEs. These simulations were performed at 300 K, using 3×3×3 supercells (with over 1200 atoms), under ambient pressure for a duration of 1 nanosecond. The figures compare the time evolution of lithium-ion MSD across different SSEs. The results indicate that at 300 K, the MSD values calculated by MatterSim are smaller than those of SevenNet and closer to the DeepMD reference values. This further confirms the superiority of MatterSim in calculating lithium-ion diffusivity in SSEs.

Overall, the study finds that MatterSim produces the most accurate results, maintaining high consistency with the DeepMD reference values across all temperatures and material types. MACE and SevenNet perform reasonably well, though their trends, while generally consistent with the reference values, still show some gaps compared to MatterSim. CHGNet and M3GNet demonstrate lower accuracy, with more pronounced deviations from the reference values at higher temperatures. ORBFF's performance is unstable, with significant deviations at high temperatures and fluctuating results in the mid-to-low temperature range. These findings underscore the importance of selecting appropriate uMLIPs for accurately calculating lithium-ion diffusivity in SSEs. MatterSim stands out with its superior performance, offering an efficient and reliable alternative for high-throughput screening of SSE materials, which could significantly reduce reliance on computationally expensive AIMD methods.

# 4. Applications

## 4.1. Structure Generation and Relaxation

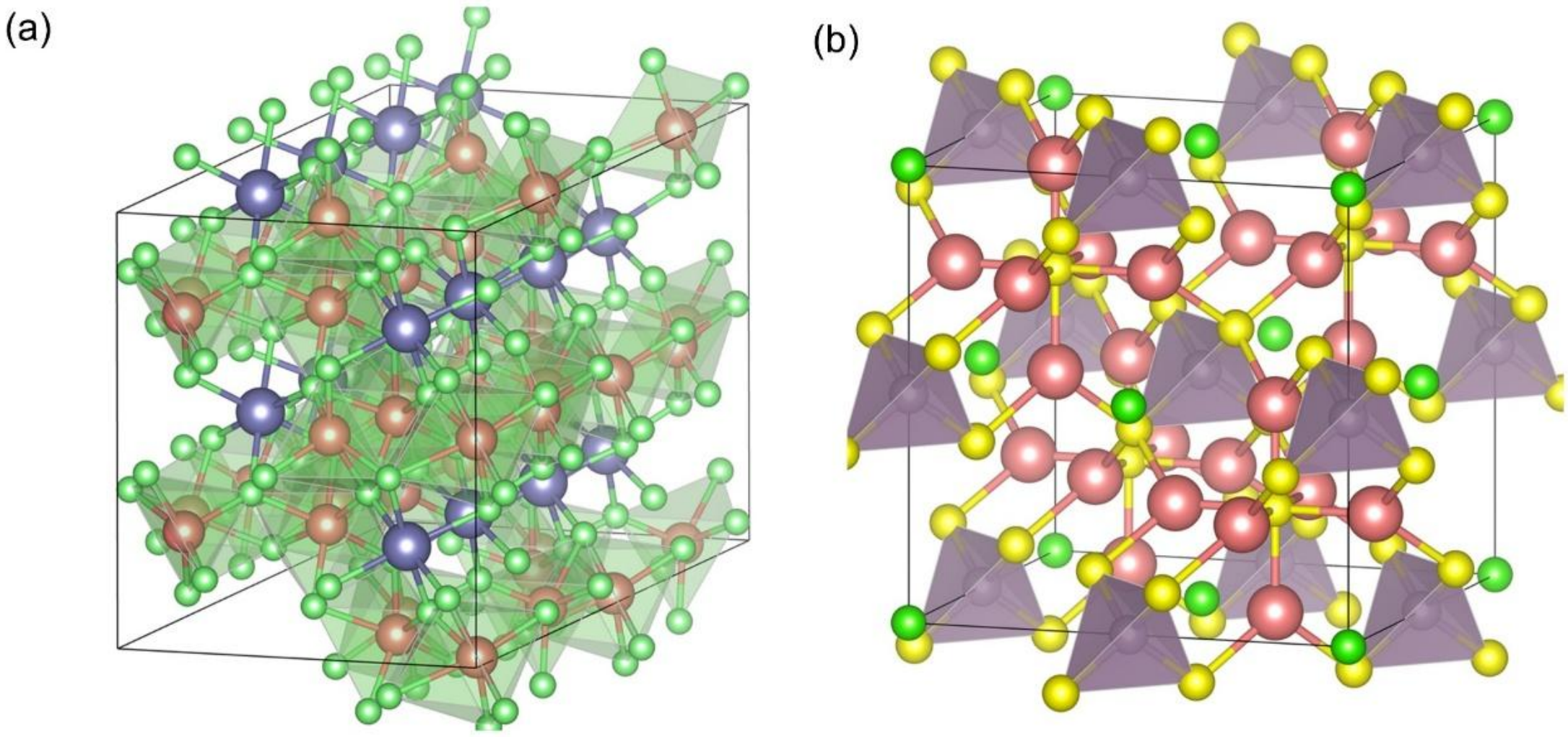

Figure 14 shows the crystal structures of $Li_3YCl_6$ and $Li_6PS_5Cl$. a presents the structure of $Li_3YCl_6$, featuring various geometries such as square pyramids, trigonal bipyramids, and octahedra, which provide multiple potential diffusivity pathways for $Li^+$ ions but also increase diffusivity complexity. b displays the lowest-energy structure of $Li_6PS_5Cl$ obtained via DFT calculations, clearly showing the distribution of $Li^+$ ions within the cage-like framework.

Through a comprehensive evaluation of uMLIPs models, including energy and force calculations, calculation of various material properties, and lithium-ion diffusivity simulations, we found that Mattersim demonstrated the best performance across all metrics. Based on this, we selected Mattersim to conduct an in-depth study of two typical SSEs—sulfide $Li_6PS_5Cl$[42] and halide $Li_3YCl_6$[43]—systematically calculating their phase structures, electrochemical stability, and ionic conductivity, while exploring the key factors influencing lithium-ion migration. Figure 14 illustrates the crystal structures of $Li_3YCl_6$ and $Li_6PS_5Cl$. The crystal structure of $Li_3YCl_6$ includes various geometric structures such as square pyramids, trigonal bipyramids, and octahedra, which provide multiple potential diffusion pathways for lithium ions. However, the relatively short bond lengths (e.g., Li(1)-Cl bond lengths of 2.44–2.67 Å), large octahedral tilting angles (42–77°), and strong binding energies significantly restrict the long-range diffusion of lithium ions. Additionally, the dynamic triangular pyramidal and planar triangular coordination among Cl atoms further complicates the diffusion process. Although $Li_3YCl_6$ possesses three-dimensional ion channels, its complex geometry and strong ion-framework interactions result in low overall ionic conductivity. $Li_6PS_5Cl$ belongs to the cubic crystal system with space group F-43m, featuring a cage-like framework composed of $S^{2-}$ and $Cl^-$ ions, which provides localized diffusion pathways for lithium ions. However, the short bond lengths between $Li^+$ and $S^{2-}$ (e.g., Li-S bond lengths of 2.33–2.42 Å) and the strong bonding limit the long-range diffusion of lithium ions. Furthermore, the rigid tetrahedral structure formed by $P^{5+}$ and $S^{2-}$ enhances the framework's stability, reducing the diffusion freedom of lithium ions. Although $SLi_6$ octahedra and $SLi_3P$ tetrahedra are connected via shared corners, offering multiple diffusion pathways, the octahedral tilting angle (approximately 54°) increases the difficulty of diffusion, resulting in lower long-range diffusion efficiency.

In summary, both $Li_3YCl_6$ and $Li_6PS_5Cl$ exhibit limited ionic conductivity due to their complex geometries, short bond lengths, and strong ion-framework interactions. Future improvements could be achieved through doping, introducing excess lithium, or optimizing structural disorder. This study not only validates the effectiveness of Mattersim in complex material systems but also provides important theoretical support for understanding the key properties of SSEs and their optimization strategies.

### 4.2. Phase, electrochemical stability and ionic conductivity

| Systems | methods | E_hull (eV/atom) | E_f (eV/atom) | Bulk Modulus (GPa) | Shear Modulus (GPa) | σ_300K (mS/cm) |
|---|---|---|---|---|---|---|
| $Li_3YCl_6$ | CHGNET | 0.0408 | -2.338 | 34.250 | 13.497 | 18.245 |
| | M3GNET | 0.0664 | -2.313 | 33.372 | 11.867 | 15.0192 |
| | ORB | 0.032 | -2.348 | 33.526 | 0.984 | 43.1116 |
| | MACE | 0.0287 | -2.351 | 40.526 | 19.903 | 3.0237 |
| | Sevennet | 0.0285 | -2.351 | 36.244 | 19.286 | 10.1981 |
| | mattersim | 0.0304 | -2.344 | 38.305 | 20.072 | 0.4553 |
| | **DFT/Deepmd** | **0.0285** | **-2.352** | **37.289** | **19.565** | 0.501 |

Table 1 presents the key performance metrics of the SSE material $Li_3YCl_6$, including formation energy(E_f), structure stability (E_above_hull), bulk modulus, shear modulus, and ionic conductivity at 300 K. These metrics were calculated using uMLIPs models and compared with reference values based on DFT to evaluate the accuracy and reliability of each model in describing material properties.

| Systems | methods | E_hull (eV/atom) | E_f (eV/atom) | Bulk Modulus (GPa) | Shear Modulus (Gpa) | σ_300K (mS/cm) |
|---|---|---|---|---|---|---|
| $Li_6PS_5Cl$ | CHGNET | 0.0774 | -1.326 | 32.36 | 13.01 | 54.0978 |
| | M3GNET | 0.0713 | -1.335 | 26.06 | 9.87 | 18.1341 |
| | ORB | 0.0678 | -1.474 | 30.87 | 0.54 | 3.5506 |
| | MACE | 0.0860 | -1.317 | 36.66 | 17.93 | 0.6692 |
| | Sevennet | 0.0821 | -1.321 | 42.35 | 22.85 | 1.3793 |
| | mattersim | 0.0835 | -1.3199 | 40.73 | 23.80 | 0.4139 |
| | **DFT/Deepmd** | **0.0830** | **-1.320** | **39.12** | **20.00** | **0.1379** |

Table 2 presents the key performance metrics of the SSE material $Li_6PS_5Cl$, including formation energy(E_f), structure stability (E_above_hull), bulk modulus, shear modulus, and ionic conductivity at 300 K. These metrics were calculated using uMLIPs and compared with reference values based on DFT to evaluate the accuracy and reliability of each model in describing material properties.

Tables 1 and 2 compare the key material property metrics of two SSE materials, $Li_6PS_5Cl$ and $Li_3YCl_6$, calculated using uMLIPs models against DFT reference values. These metrics include energy stability, formation energy, bulk modulus, shear modulus, and room-temperature ionic conductivity ($\sigma_{300}$K). Overall, MatterSim performs best, particularly excelling in the calculation of lithium-ion conductivity, with results significantly closer to DeepMD and within the same order of magnitude, demonstrating its superiority in evaluating SSE properties. SevenNet and MACE follow as secondary performers, while CHGNet and M3GNet exhibit lower accuracy. ORBFF performs the worst and shows instability. Supplementary Figures S48 and S49 further validate the reliability of MatterSim: Figure S48 compares the energy-volume relationship, and Figure S49 contrasts the lithium-ion conductivity. The results show that MatterSim is highly consistent with reference values in both energy-volume relationships and conductivity calculation, providing crucial theoretical support for designing novel SSEs while reducing reliance on traditional DFT and MD calculations.

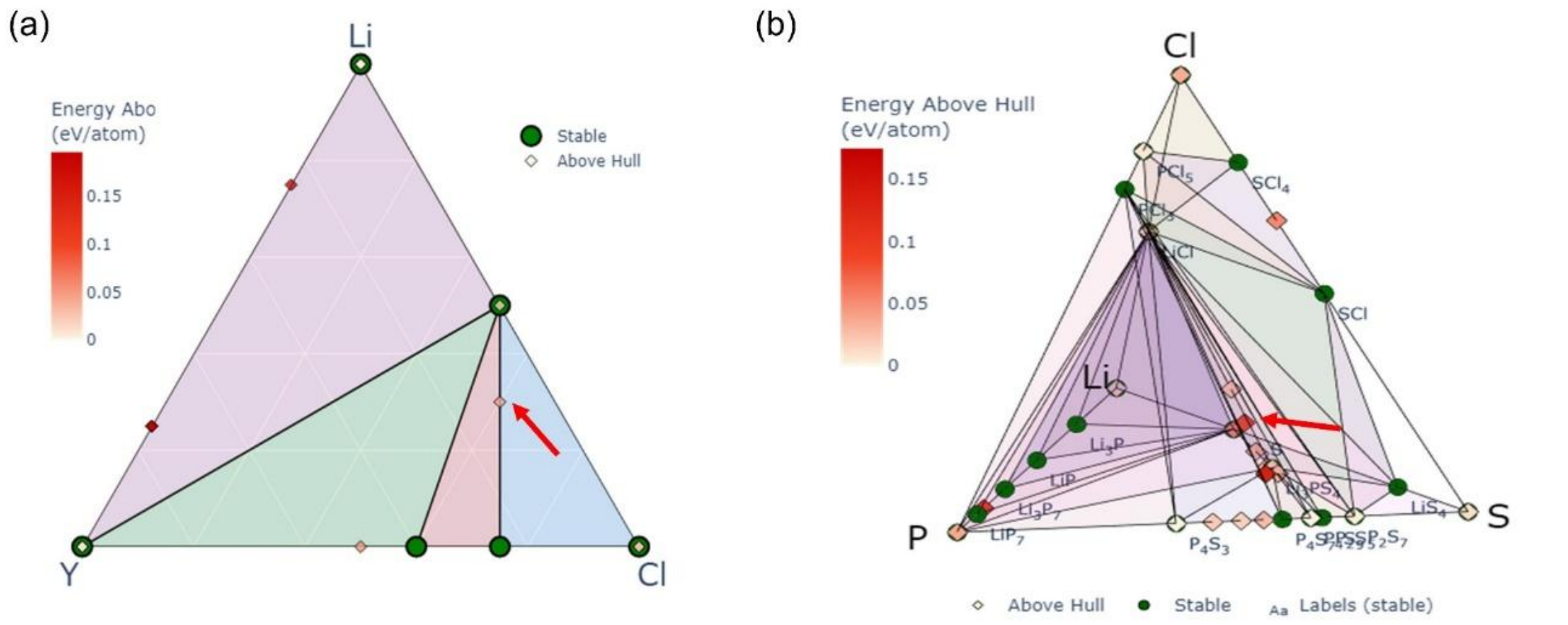

Figure 15 illustrates the energy and structural composition of $Li_3YCl_6$ and $Li_6PS_5Cl$ calculated using MatterSim, combined with an evaluation of their thermal stability using Pymatgen within the Li-Y-Cl ternary phase diagram and the Li-P-S-Cl quaternary phase diagram. a shows the position of $Li_3YCl_6$ in the ternary phase diagram (marked by a red arrow), with an E_above_hull value of 0.0304, which is close to the DFT-calculated value of 0.0285. b presents the position of $Li_6PS_5Cl$ in the quaternary phase diagram, with an E_above_hull value of 0.0835, highly consistent with the DFT value of 0.0830. These results demonstrate that this method can effectively assess the thermal stability of materials with high computational accuracy and reliability.

To evaluate the phase stability of $Li_3YCl_6$ and $Li_6PS_5Cl$, we constructed their 0 K phase diagrams within their respective chemical spaces (Li−Y−Cl and Li−P−S−Cl) in the Figure 15. First, we used MatterSim combined with the Atomic Simulation Environment (ASE) tools to optimize the structures and calculate their energies. Subsequently, Pymatgen was utilized to obtain all possible combination phases within the chemical spaces (e.g., Li, Y, Cl, and their combinations for $Li_3YCl_6$; Li, P, S, Cl, and their combinations for $Li_6PS_5Cl$). Based on this data, calculations were performed on three-dimensional (for $Li_3YCl_6$) or four-dimensional (for $Li_6PS_5Cl$) convex hulls, constructed based on normalized energy per atom and elemental atomic fractions. The vertices of the convex hulls were projected onto the composition coordinate space to generate the phase diagrams. Next, the E_above_hull values (energy differences relative to stable phases) of $Li_3YCl_6$ and $Li_6PS_5Cl$ were calculated. The results show that the E_above_hull for $Li_3YCl_6$ is 30.4 meV/atom (DFT reference value: 28.5 meV/atom), and for $Li_6PS_5Cl$, it is 83.5 meV/atom (DFT reference value: 83.0 meV/atom). Although both materials are metastable at 0 K (e.g., $Li_3YCl_6$ relative to $YCl_3$ and LiCl, and $Li_6PS_5Cl$ relative to $Li_3PS_4$, $Li_2S$, and LiCl), they still exhibit a certain degree of stability under practical conditions. In summary, the MatterSim method comprehensively evaluates the phase stability of these two materials and analyzes their thermodynamic stability based on E_above_hull, demonstrating their potential application value.

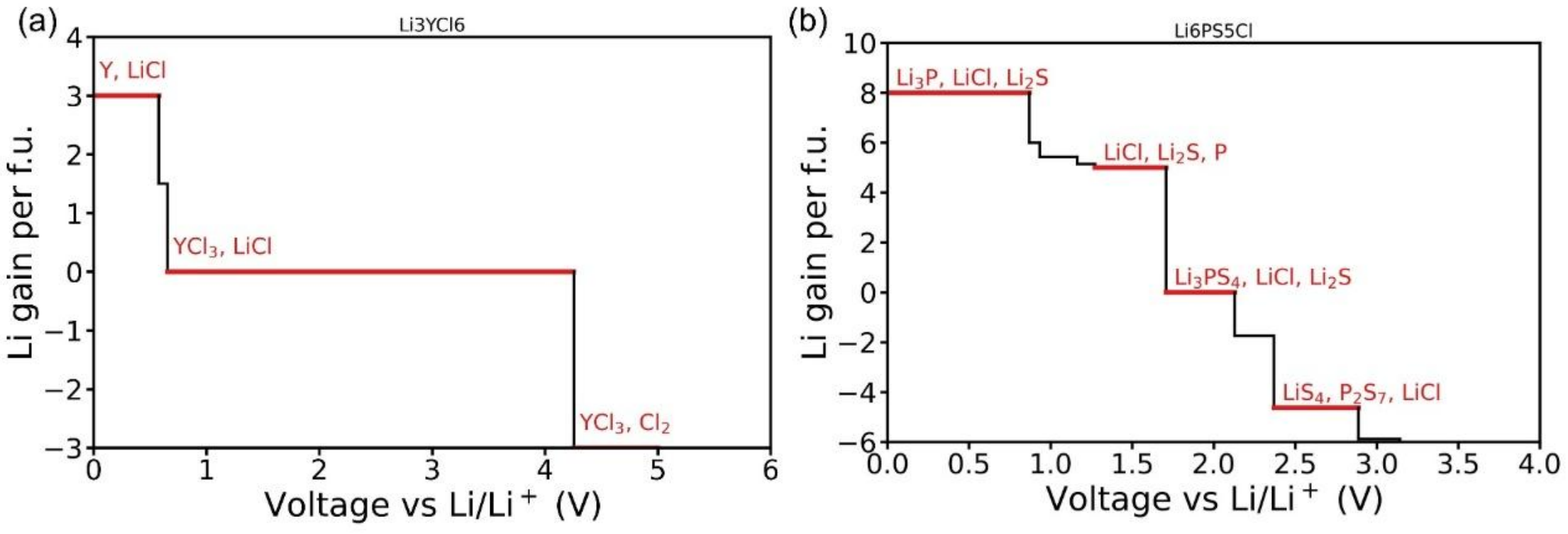

Figure 16, a and b respectively show the variation in $Li^+$ absorption per formula unit for $Li_3YCl_6$ and $Li_6PS_5Cl$ at different voltages (relative to $Li/Li^+$). At low voltages (high Li chemical potential), both materials undergo reduction and absorb $Li^+$; at high voltages (low Li chemical potential), they undergo oxidation and release $Li^+$. The calculated phase equilibria for selected voltage ranges are annotated in the figure for simplicity.

Figure 16 illustrates the relationship between the amount of $Li^+$ absorption per formula unit and voltage (relative to $Li/Li^+$) for $Li_3YCl_6$ (a) and $Li_6PS_5Cl$ (b) at different voltages. In this study, MatterSim combined with ASE tools was used to precisely calculate the material energies, and their electrochemical stability was evaluated within their respective chemical spaces (Li−Y−Cl and Li−P−S−Cl). The results indicate that at low voltages (high Li chemical potential), both materials undergo reduction reactions and absorb $Li^+$, while at high voltages (low Li chemical potential), they undergo oxidation reactions and release $Li^+$. To more accurately assess stability, we adopted a lithium chemical potential reference method based on bulk lithium energy rather than the traditional internal energy phase diagram approach. This method effectively reveals the electrochemical behavior of the materials across different voltage ranges. $Li_3YCl_6$ exhibits a broad electrochemical stability window (0.6 to 4.2 V), indicating its stability across a wide range of battery operating voltages, which is critical for safe battery operation. In contrast, $Li_6PS_5Cl$ has a narrower stability range (1.7 to 2.3 V). However, under extreme voltages, $Li_6PS_5Cl$ forms electronically insulating but ionically conductive phases (e.g., $Li_2S$ and LiCl), which act as passivation layers to prevent further decomposition, thereby protecting the electrolyte and extending battery life. Notably, the calculation results from MatterSim are highly consistent with DFT, with calculated electrochemical stability windows and key phase transition points closely matching DFT reference values, fully validating its high accuracy. Through the above analysis, MatterSim not only provides precise energy data but also demonstrates significant application potential in the study of halide and sulfide electrolytes.

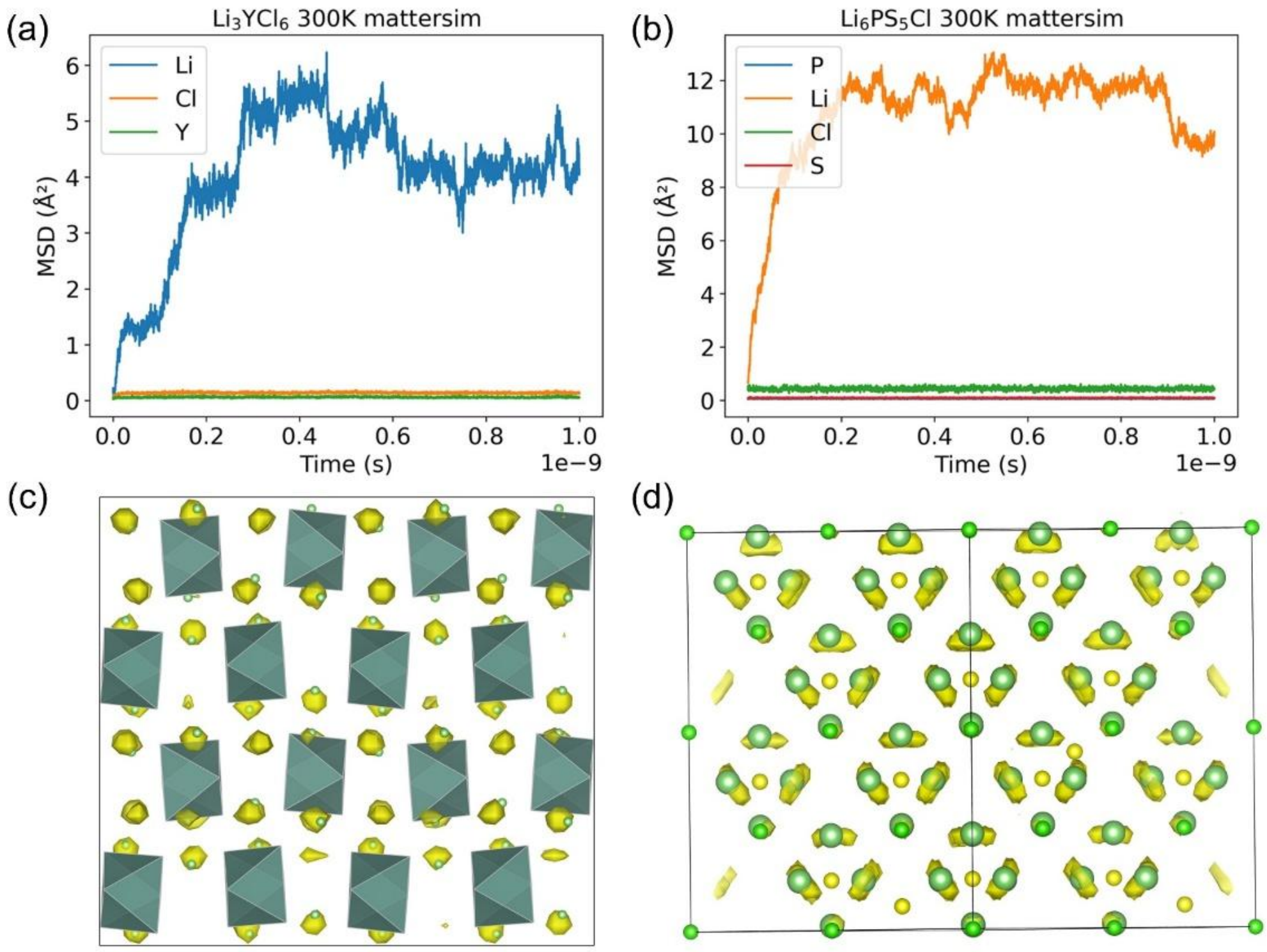

Figures 17, a and b show the MSD of lithium ions in $Li_3YCl_6$ and $Li_6PS_5Cl$ over time at 300 K, as simulated by MatterSim for 1 ns. These results intuitively reflect the diffusion dynamics of lithium ions in the two materials. c and d present the probability density distribution of lithium ions under high isosurfaces, with yellow isosurfaces representing the probability density along the diffusion pathways, drawn based on an isovalue of $0.002/a_0^3$ (where $a_0$ is the Bohr radius).

Through the above studies, we found that MatterSim performed best in evaluating lithium-ion diffusion properties across 18 systems, including halides, oxides, and sulfides. Based on this, we further utilized MatterSim to conduct MD simulations of $Li_3YCl_6$ and $Li_6PS_5Cl$ under ambient pressure, using a 3×3×3 supercell (with more than 1200 atoms) at different temperatures, to investigate the diffusion behavior of lithium ions in depth. At 300 K, we performed 1 ns of NVT MD simulations, extracted diffusion rate data, and analyzed the relationship between mean square displacement (MSD) and time. Using the Arrhenius relation and the Nernst-Einstein equation, we derived the ionic conductivity at room temperature. Figures 17a and b show the MSD of lithium ions in $Li_3YCl_6$ and $Li_6PS_5Cl$ over time. The results indicate that the MSD of lithium ions in both materials stabilizes after a few hundred picoseconds, suggesting that their diffusion is significantly restricted, which is unfavorable for long-range transport. Figures 17c and d respectively display the probability density distribution of lithium ions in the two materials, with yellow isosurfaces reflecting the diffusion paths and probability densities of lithium ions.

For $Li_3YCl_6$ (Figure 17c), the coordination environment of lithium ions includes various geometries such as square pyramids, trigonal bipyramids, and octahedra. These complex structures provide multiple potential diffusion pathways for lithium ions but also increase the complexity of diffusion. For example, the square pyramid formed by the Li(1) site shares corners or edges with multiple octahedra and trigonal bipyramids, potentially offering multi-path diffusion. However, the connections between these paths require overcoming high energy barriers. Additionally, the short bond lengths between lithium ions and Cl atoms (e.g., Li(1)-Cl(1) at 2.67 Å and Li(1)-Cl(2) at 2.44 Å) indicate strong bonding, further restricting the diffusion of lithium ions. Particularly, the range of tilt angles for corner-sharing octahedra (42–77°) increases the difficulty for lithium ions to hop between octahedra.

For $Li_6PS_5Cl$ (Figure 17d), lithium ions tend to move within a cage-like environment surrounding the S 4d sites, forming distorted $Li_6S$ octahedra. The high-value isosurfaces reveal that the motion of lithium ions is mainly confined within specific cage-like structures, and this localized diffusion contributes little to long-range transport. The cage-like structure, composed of S and Cl atoms, allows lithium ions to vibrate rapidly but rarely cross into other cages, further limiting long-range ion transport. This localized behavior suggests that although $Li_6PS_5Cl$ exhibits some degree of local ion diffusion capability, its overall ionic conductivity remains constrained by the cage-like structure. To further explore the influence of temperature on the diffusion behavior of lithium ions, we conducted 100 ps MD simulations of $Li_3YCl_6$ and $Li_6PS_5Cl$ at higher temperatures (700 K and 1100 K). Figures S50a and b illustrate the MSD of $Li_3YCl_6$ at 700 K and 1100 K, while Figures S51a and b present the MSD of $Li_6PS_5Cl$ at the same temperatures. These results visually reflect the diffusion dynamics of lithium ions in the two materials under high-temperature conditions. As the temperature increases, the MSD of lithium ions significantly rises, indicating that thermal activation promotes diffusion. However, even at high temperatures, the diffusion of lithium ions still exhibits a certain degree of localization. Meanwhile, Figures S50c and d, as well as Figures S51c and d, respectively display the probability density distribution of lithium ions in $Li_3YCl_6$ and $Li_6PS_5Cl$ at 700 K and 1100 K. The yellow isosurfaces in the figures represent the probability density of lithium ions along the diffusion paths, plotted based on the $0.002/a_0^3$ contour line (where $a_0$ is the Bohr radius). The high-value isosurfaces reveal the short-range diffusion behavior of lithium ions within local cage-like structures, while the low-value isosurfaces more clearly show the long-range diffusion paths of lithium ions between different cages. These analyses indicate that although higher temperatures enhance the diffusion rate of lithium ions, the ion transport in both materials remains strongly constrained by local structures.

In summary, this study reveals the localized characteristics of lithium-ion diffusion in $Li_3YCl_6$ and $Li_6PS_5Cl$ through MatterSim simulations and their limitations on long-range diffusion, providing important insights into the ion transport mechanisms of these two SSEs. It also points out directions for improving their ionic conductivity through structural optimization.

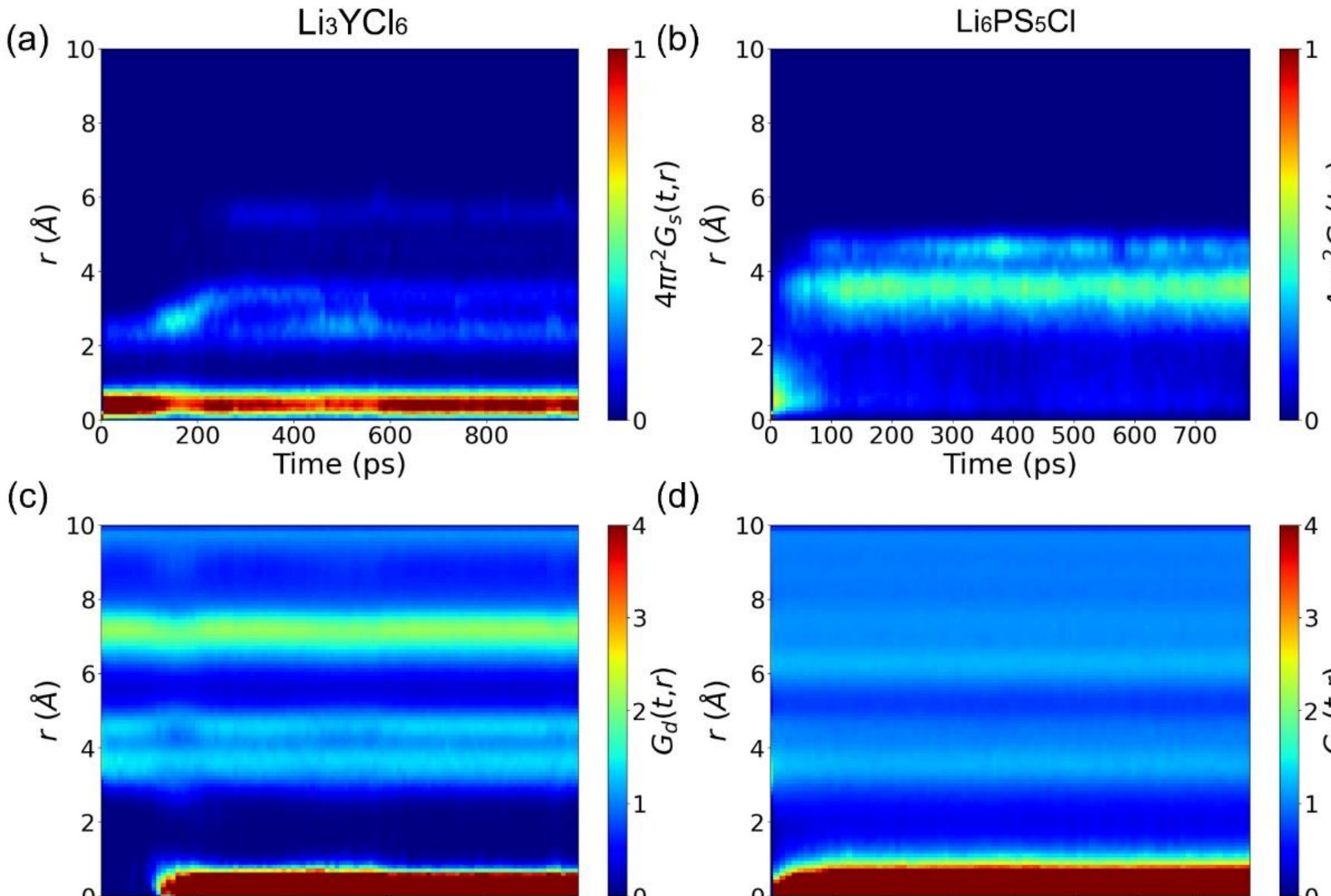

Figure 18 shows the van Hove correlation functions of $Li_3YCl_6$ and $Li_6PS_5Cl$ at ambient pressure and 300 K, obtained through MatterSim simulations. The upper panels display the self-part (Gs), reflecting local lithium-ion motion, while the lower panels show the distinct part (Gd), revealing long-range diffusion characteristics. Specifically, a and c represent Gs and Gd for $Li_3YCl_6$, while b and d correspond to Gs and Gd for $Li_6PS_5Cl$.

The van Hove correlation function is a crucial tool for describing the spatial and temporal distribution of particles, divided into the self-part (Gs) and distinct part (Gd). Gs(r,t) represents the probability distribution of particles diffusing from their initial positions, reflecting local diffusion behavior, with peak positions indicating regions of high probability. Gd(r,t) describes the radial distribution of other particles relative to a reference particle, revealing long-range diffusion behavior, with peaks also corresponding to high-probability regions. By analyzing the van Hove function, we can gain deeper insights into the diffusion paths and dynamic behavior of lithium ions. Figures 18a and c show the Gs and Gd results for $Li_3YCl_6$. The Gs plot reveals a persistent peak in the 2 to 4 Å range, with additional peaks at larger r values appearing after approximately 300 ps. This indicates that lithium ions in $Li_3YCl_6$ not only diffuse within cages but also migrate between cages, achieving a degree of long-range diffusion. The Gd plot further shows that $Li_3YCl_6$ exhibits a broader high-probability distribution over shorter timescales, with more high-probability peaks emerging over time, suggesting frequent cage-to-cage transitions and effective long-range diffusion. Figure18b and d present the Gs and Gd results for $Li_6PS_5Cl$. The Gs plot shows a persistent peak in the 4 to 5 Å range, corresponding to the nearest Li-Li distances within the same cage, indicating that lithium ions are primarily confined to intracage diffusion with limited contribution to long-range diffusion. The Gd plot confirms this localized behavior, showing that high-probability distributions are concentrated within short-time intracage diffusion, with minimal intercage transitions. Figure 18 complements the MSD results over time calculated by MatterSim in Figure 17, jointly validating the accuracy and reliability of the simulations. Through the analysis of the van Hove correlation function, we not only uncover the diffusion paths and dynamic behavior of lithium ions in $Li_3YCl_6$ and $Li_6PS_5Cl$ but also provide important theoretical insights into their ion transport mechanisms.

### 4.3. Exploration of factors affecting ionic conductivities

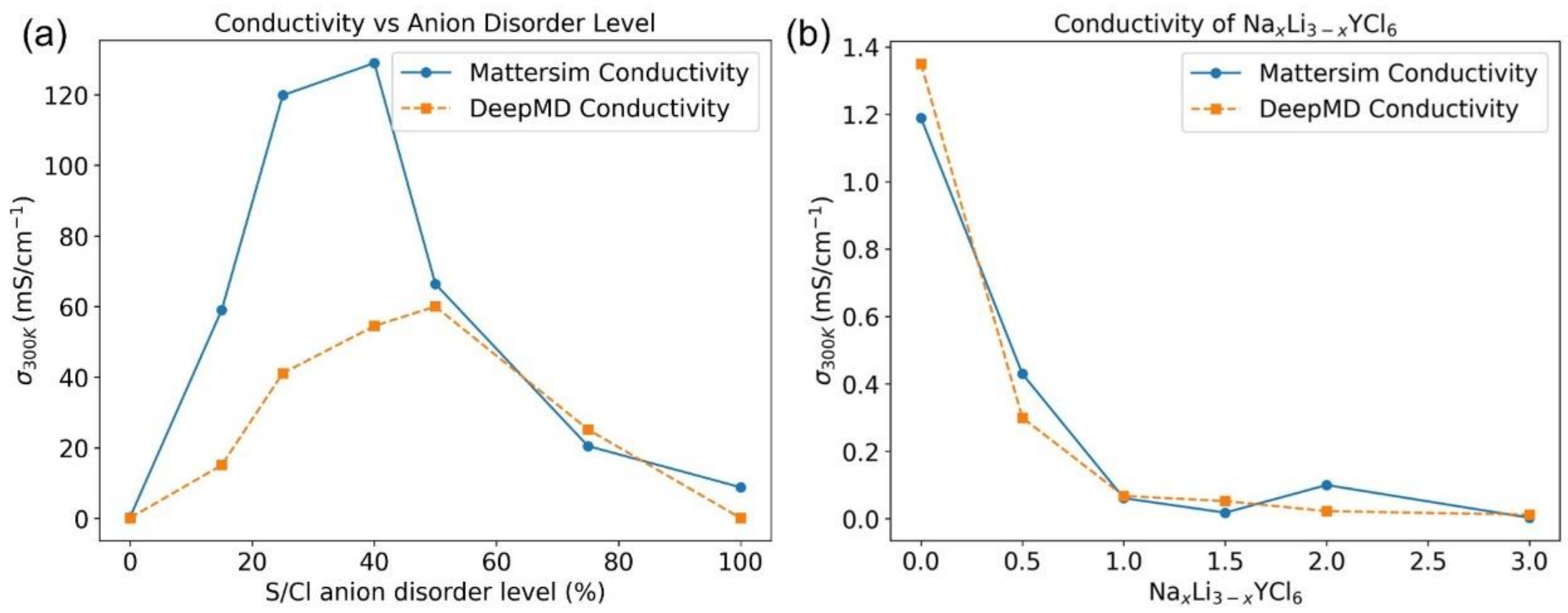

Figure 19a shows the variation in lithium-ion conductivity of $Li_6PS_5Cl$ at room temperature with different levels of S/Cl anion disorder. Except for 0% and 100% disorder, each data point is obtained by averaging the ionic conductivity calculated from multiple structures of different S/Cl arrangements using MD simulations based on the MatterSim model. Figure 19b illustrates the trend in ionic conductivity of the $NaxLi_{3-x}YCl_6$ series at room temperature as the Li decreases, demonstrating that the introduction of Li significantly enhances the material's ion transport capability.

This study investigates the lithium-ion diffusion mechanism in $Li_6PS_5Cl$ and its relationship with S/Cl anion disorder levels using MD simulations based on the MatterSim model. By exchanging Cl at the 4a site with S at the 4d site, $Li_6PS_5Cl$ structures with varying S/Cl disorder levels (15%, 25%, 40%, 50%, and 75%) were generated. For 0% and 100% disorder levels, only one structure was considered and optimized using DFT. For other disorder levels, the four structures with the lowest Ewald energies were selected for DFT optimization, retaining the most stable configuration. During the MatterSim optimization process, the SCF energy convergence threshold was set to $10^{-5}$ eV, and the force convergence threshold was set to 0.05 eV/Å.

Figure 19a illustrates the effect of different S/Cl disorder levels on lithium-ion conductivity at room temperature. The results show that the lithium-ion conductivity initially increases with increasing disorder levels, reaching a peak at approximately 40%-50%, before decreasing. This trend is consistent with DeepMD calculations, and the similar MSD variation trend is also shown in Supplementary Figure S52. Figure S54 compares the MSD and lithium-ion probability density distributions at 0% and 40% disorder levels, further validating the accuracy of the MatterSim calculations. The analysis indicates that low disorder levels result in discontinuous inter-cage hopping pathways, leading to low conductivity. As the disorder level increases, effective hopping pathways form, enhancing conductivity. However, when the disorder level exceeds 40%, dual-ion hopping pathways are suppressed, causing the diffusion network to disconnect again and resulting in reduced conductivity. These findings reveal the influence of S/Cl disorder levels on the connectivity of lithium-ion diffusion pathways, providing theoretical guidance for optimizing SSE performance.

Figure 19b shows the ionic conductivity of NaxLi$_{3-x}$YCl$_6$ at room temperature, indicating a significant increase in conductivity with increasing Li content. This result aligns with DeepMD calculations and the MSD trends presented in Supplementary Figure S53. Figure S55 compares the MSD and lithium-ion probability density distributions of $Li_3YCl_6$ and $Na_{1.5}Li_{1.5}YCl_6$, further verifying the reliability of MatterSim. The smaller radius (0.76 Å) and high mobility of $Li^+$ ions are key factors contributing to the enhanced conductivity. Additionally, increasing Li content may optimize the crystal structure by expanding diffusion channels or introducing local disorder, thereby reducing migration energy barriers. For example, the activation energy of $Na_2LiYCl_6$ (x = 2) is 0.6 eV, lower than the 0.82 eV of $Na_3YCl_6$ (x = 3), indicating that increased Li content significantly improves low-temperature ion migration capability[43]. Higher Li content also increases the concentration of mobile ions, optimizing charge balance and local electric field distribution, further promoting ion transport. Collectively, these factors enhance the material's ionic conductivity, confirming the potential of NaxLi$_{3-x}$YCl$_6$ as a high-performance SSE and providing important insights for designing novel ion-conductive materials.

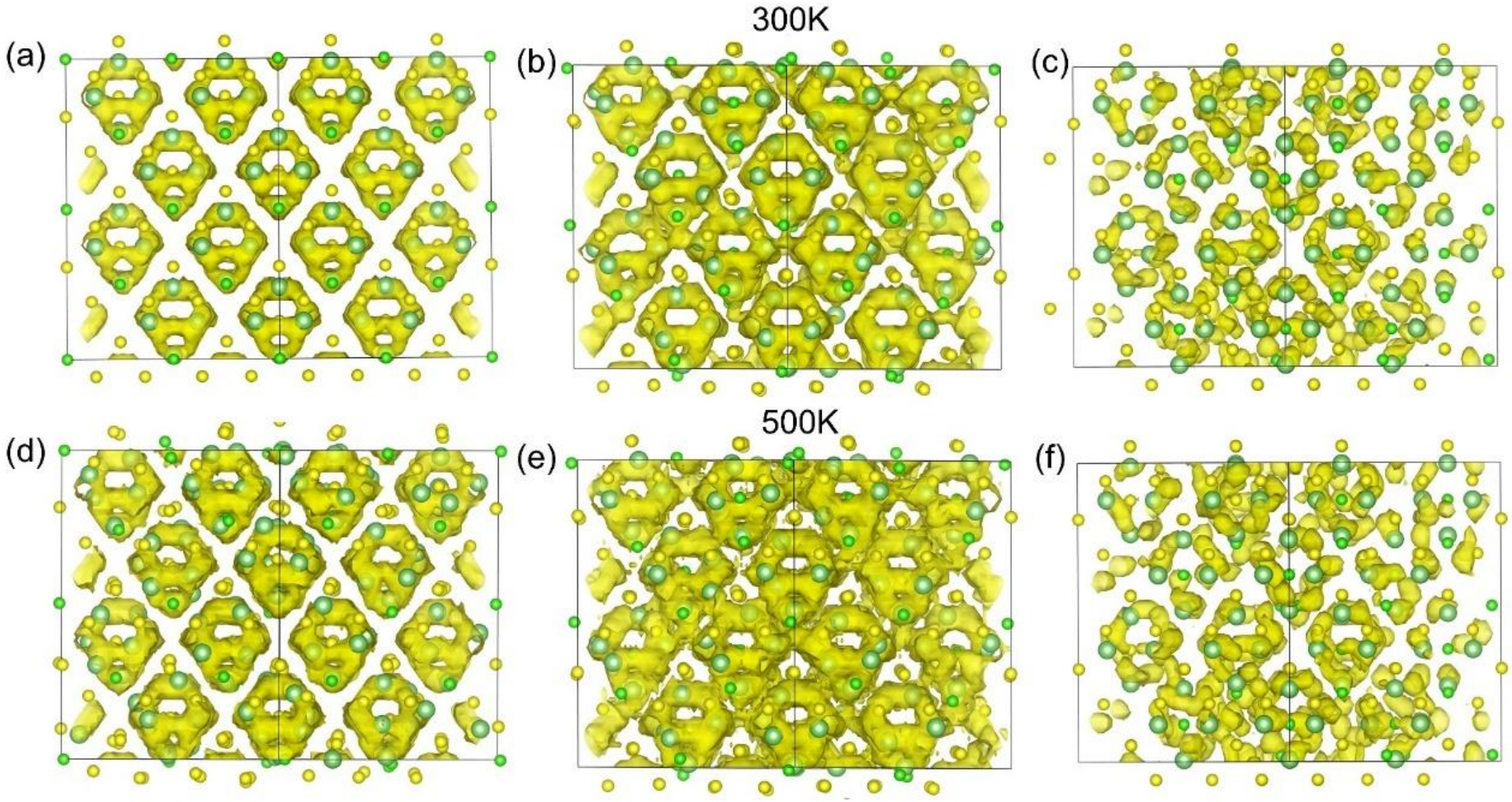

Figure 20 shows the lithium-ion probability density distributions obtained via MD simulations using MatterSim at 300 K and 500 K, for different S/Cl anion disorder levels (0%, 40%, and 75%). The isosurfaces are plotted based on a contour level of $0.0002/a_0^3$ (where $a_0$ is the Bohr radius).

Figure 20 illustrates the probability density distribution of lithium ions at different S/Cl anion disorder levels, obtained through MD simulations at 300 K and 500 K. At 0% S/Cl disorder level, lithium ions are confined within isolated lithium clusters, with a lack of significant connections between clusters, leading to restricted lithium-ion diffusion and consequently lower ionic conductivity. At 40% S/Cl disorder level, continuous diffusion pathways form between lithium clusters, allowing lithium ions to move more freely, resulting in higher ionic conductivity. However, at 75% S/Cl disorder level, lithium ions are again localized in specific regions, and the diffusion pathways become discontinuous, causing a decline in lithium-ion conductivity. These changes in lithium-ion probability density distribution visually reveal the influence of S/Cl anion disorder levels on lithium-ion diffusion behavior, further explaining why $Li_6PS_5Cl$ exhibits higher lithium-ion conductivity around 40% S/Cl disorder levels. This result indicates that an appropriate S/Cl disorder level can optimize the connectivity of lithium-ion diffusion pathways, thereby enhancing the overall ionic transport performance of the material.

| Systems | Ionic conductivities and | $Na_xLi_{3-x}YCl_6$ | Energy (eV) | σ300K (mS/cm) | σ300K (mS/cm) | Li ions diffusiviy | Na ions diffusivit |
|---|---|---|---|---|---|---|---|

| | relaxed calculator | compositions | | Li ionic conductivities | Na ionic conductivities | 300K D($m^2$/s) | y 300K D($m^2$/s) |
|---|---|---|---|---|---|---|---|
| $Na_xLi_{3-x}YCl_6$ 1080 atoms 3*3*3 super cell, NVT ensemble, 1ns | mattersim | $Na_{0.5}Li_{2.5}YCl_6$ | -171.428 | 8.988 | 0.0780 | 1.36e-11 | 5.92e-13 |
| | | | -171.352 | 0.245 | 0.0036 | 3.5e-13 | 2.63e-14 |
| | | | -171.305 | 0.430 | 0.363 | 6.40e-13 | 2.64e-12 |
| | | | -171.308 | 4.488 | 0.0016 | 6.53e-12 | 1.19e-14 |
| | | $Na_{1.0}Li_{2.0}YCl_6$ | -170.283 | 0.962 | 0.241 | 1.90e-12 | 9.54e-13 |
| | | | -169.910 | 0.127 | 0.206 | 2.32e-13 | 7.52e-13 |
| | | | -169.856 | 1.655 | 0.623 | 3.01e-12 | 2.27e-12 |
| | | | -169.920 | 0.061 | 0.058 | 1.11e-13 | 2.12e-13 |
| | | | -169.977 | 0.0724 | 0.270 | 1.31e-13 | 9.83e-13 |
| | | $Na_{1.5}Li_{1.5}YCl_6$ | -169.148 | 0.3009 | 0.0279 | 8.19e-13 | 7.61e-14 |
| | | | -168.433 | 0.220 | 0.00397 | 5.33e-13 | 9.66e-15 |
| | | | -168.364 | 0.018 | 1.242 | 4.38e-14 | 3.02e-12 |
| | | | -168.398 | 1.562 | 0.071 | 3.79e-12 | 1.70e-13 |
| | | $Na_{2.0}Li_{1.0}YCl_6$ | -168.1178 | 2.508 | 0.936 | 1.06e-11 | 1.97e-12 |
| | | | -168.227 | 0.189 | 0.029 | 8.04e-13 | 6.26e-14 |
| | | | -168.106 | 4.911 | 2.663 | 2.07e-11 | 5.62e-12 |
| | | | -168.161 | 0.268 | 0.208 | 1.13e-12 | 4.40e-13 |

Table 3 presents the results of MD simulations using MatterSim at 300 K, with a 3×3×3 supercell and NVT ensemble, for $NaxLi_{3-x}YCl_6$ materials with different x values (x = 0.5, 1.0, 1.5, 2.0). The simulation time was 1 ns, and the ionic conductivities and diffusion rates of lithium and sodium ions were calculated for various Na/Li arrangements. These results provide crucial data support for understanding the relationship between ion transport behavior and Na/Li arrangements in $NaxLi_{3-x}YCl_6$.

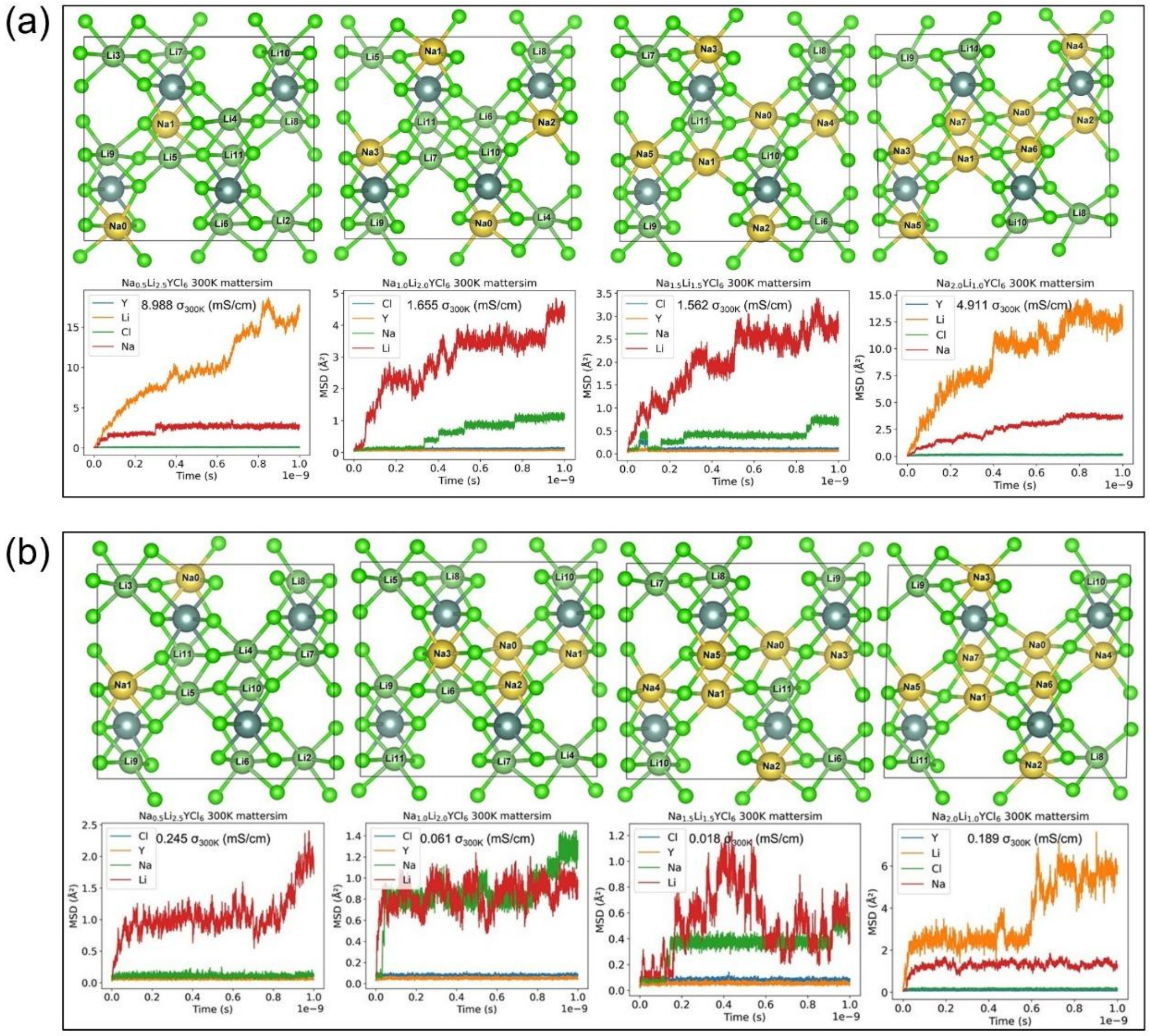

Figure 21 shows the impact of Na/Li atomic arrangements on ionic conductivity in $NaxLi_{3-x}YCl_6$ structures with different x values (x = 0.5, 1.0, 1.5, 2.0) using MatterSim. a and b correspond to the Na/Li arrangements for the maximum and minimum lithium-ion conductivity in Table 3, respectively, with comparisons of the MSD over time under the respective conditions. These comparisons intuitively reveal the critical role of Na/Li arrangements in ion transport behavior.

In studying $NaxLi_{3-x}YCl_6$, we combined MatterSim with the Pymatgen library to address the issue of solid-solution disorder in Na/Li positions within the crystal structure and generated ordered structures. The specific steps were as follows: The prototype structure of $Li_3YCl_6$ was obtained from the ICSD database. By partially replacing Li with Na (x = 0.5, 1.0, 1.5, 2.0), all symmetrically distinct structures were enumerated using Pymatgen's enumlib tool, followed by geometric optimization using MatterSim. The lowest-energy structures were ultimately selected (see Table 3) to provide a foundation for subsequent property calculations. Based on the MatterSim, NVT MD simulations were conducted for 1 ns under ambient pressure, using a 3×3×3 supercell (with more than 1200 atoms) at 300 K, to calculate the ionic conductivity and diffusivity of lithium and sodium ions at different solid-solution ratios. The results indicate: (1) The introduction of sodium ions significantly reduces ionic conductivity, with sodium conductivity being much lower than that of lithium ions; (2) Different Na/Li arrangements significantly affect ionic conductivity, with notable differences observed among structures with the same solid-solution ratio.

Figure 21 further analyzes the impact of Na/Li arrangement on ionic conductivity. Figure 21a shows the Na/Li arrangement corresponding to the maximum lithium-ion conductivity for different x values (x = 0.5, 1.0, 1.5, 2.0) in Table 3, along with the time evolution of MSD; Figure 21b presents the corresponding results for the minimum conductivity. Comparative analysis reveals that high-conductivity structures typically exhibit optimized coordination network topology and continuous three-dimensional diffusion pathways. For example, in $Na_{0.5}Li_{2.5}YCl_6$, the high-conductivity structure features $NaCl_6$ octahedra densely interconnected via edge- and face-sharing with $LiCl_6$ and $YCl_6$ octahedra, significantly reducing the migration energy barrier for lithium ions. In contrast, in low-conductivity structures, the connection patterns of $NaCl_6$ octahedra are weakened, leading to broken channels and localized rigidity, which suppresses lithium-ion diffusion. Similarly, in $Na_{1.0}Li_{2.0}YCl_6$, the high-conductivity structure (orthorhombic $Pna2_1$) constructs efficient migration pathways through uniform bond length distribution and continuous vertex, edge, and face sharing, whereas the low-conductivity structure (triclinic P1) suffers from complex low symmetry and fragmented local coordination environments, hindering cooperative diffusion. For $Na_{1.5}Li_{1.5}YCl_6$, the monoclinic $P2_1$ structure enhances lithium-ion diffusion efficiency by simplifying cation distribution and optimizing edge/face sharing, while the triclinic P1 structure forms high-energy barrier bottlenecks due to distorted octahedra and complex Cl coordination modes. In $Na_{2.0}Li_{1.0}YCl_6$, the high-conductivity structure achieves efficient diffusion through an open framework and low activation energy pathways, whereas the low-conductivity structure experiences broken pathways and stress accumulation due to extreme bond length variations and local lattice distortions, further inhibiting ion transport.

These results demonstrate that differences in lithium-ion conductivity primarily arise from the combined effects of crystal structure symmetry, coordination network connectivity, and local lattice distortions. They reveal the key mechanisms for optimizing Na/Li arrangements to enhance ionic conductivity, providing important theoretical guidance for designing high-performance SSE materials.

## 5. Discussion

With the rapid development of uMLIPs models, how to comprehensively and accurately evaluate their performance has become a key challenge in the field. Our framework, through multi-level and multi-dimensional test designs, covers core metrics such as energy, forces, thermodynamic properties, elastic moduli, and lithium-ion diffusion behavior, ensuring a thorough examination of model performance while highlighting the critical roles of training data quality and physical consistency in model performance. This provides a standardized tool for systematically evaluating the performance of uMLIPs models and lays a solid foundation for the development and optimization of novel uMLIPs models in the future. Through systematic evaluation of uMLIPs models, we found that MatterSim excels in almost all assessment metrics, particularly demonstrating significantly superior predictive accuracy and generalization capabilities in complex material systems compared to other models. Its outstanding performance is attributed to advanced architecture design and the application of active learning strategies, enabling it to efficiently capture many-body interactions and exhibit excellent robustness across chemical spaces. In contrast, models like ORBFF and CHGNet, while performing adequately in structure optimization and static energy calculation, show significant deviations in dynamic behavior and thermodynamic property calculation, especially under extreme conditions. This further underscores the critical impact of model design philosophy (e.g., energy derivative consistency and training data coverage) on prediction accuracy and generalization capability.

Based on MatterSim simulations, we conducted an in-depth analysis of the ion transport mechanisms and key influencing factors of two typical SSEs, $Li_6PS_5Cl$ and $Li_3YCl_6$. The study reveals that S/Cl anion disorder levels and Na/Li arrangements significantly affect the connectivity of lithium-ion diffusion pathways. For instance, in $Li_6PS_5Cl$, an S/Cl disorder level of approximately 40%-50% can form a continuous diffusion network, significantly enhancing ionic conductivity; however, excessively high disorder levels disrupt pathway connectivity, leading to a decline in conductivity. Similarly, in the $Na_xLi_{3-x}YCl_6$ system, increasing lithium content expands diffusion channels and introduces local disorder, lowering migration energy barriers and thereby improving ionic conductivity. These findings provide theoretical guidance for enhancing SSE performance through structural optimization and reveal the combined influence of crystal structure, coordination networks, and local distortions on ion transport behavior.

Looking ahead, the potential applications of uMLIPs models in materials science are vast, but some key challenges remain to be addressed. First, improving the computational capabilities of models under extreme conditions (e.g., high pressure, high temperature, or electrochemical cycling) is crucial for achieving their widespread application. Second, integrating multi-scale modeling methods (e.g., coupling molecular dynamics with continuum mechanics) can provide a more comprehensive description of the actual dynamic behavior of SSEs. Additionally, by incorporating more diverse training data (e.g., systems containing rare earth elements and complex crystal structures), the generalization ability of models in unseen chemical spaces can be enhanced, supporting the discovery and optimization of new materials. Ultimately, with the continuous improvement of machine learning tools, uMLIPs will play a pivotal role in the development of next-generation energy materials, laying the groundwork for high-performance energy storage technologies.

## Acknowledgements

We are grateful for the financial support from the National Key Research and Development Program of China (Grant Nos.2021YFB3702104). The computations in this paper were run on the π 2.0 cluster supported by the Center for High Performance Computing at Shanghai Jiao Tong University.